\documentclass[10pt,journal,twocolumn,twoside]{IEEEtran}
\usepackage{epsfig,amsfonts,color,amsbsy}
\usepackage[nolist]{acronym}
\usepackage{graphicx,cite,amssymb,bm}
\usepackage{amsmath}
\usepackage{mathrsfs}
\usepackage{bm}
\usepackage{multirow}
\usepackage{bigstrut}
\usepackage{psfrag}
\usepackage{stackengine}
\usepackage{xcolor}
\usepackage{mathtools}
\usepackage{tabularx,array}
\usepackage{mathbbol}
\usepackage{lipsum}
\usepackage[sans]{dsfont}
\usepackage{array, makecell}
\usepackage{float}
\usepackage{verbatim}
\usepackage{mathrsfs}
\usepackage{pgfplots}
\usepackage{scalerel}
\usepackage{mathtools}
\usepackage{derivative}
\usepackage{amsthm}
\usepackage[caption=false,font=footnotesize]{subfig}
\usepackage{flushend}
\usepackage{url}
\usepackage{siunitx}
\usepackage{multirow}
\usepackage{graphicx}
\usepackage{pgfplots}
\usepackage{bm}
\usepackage{paralist}
\usepackage{xspace}
\usepackage{fancyref}
\usepackage{bbm}
\usepackage{tikz}
\usepackage{pgfplots}
\usetikzlibrary{shapes,arrows}
\usetikzlibrary{calc}

\definecolor{TUMBlack}{cmyk}{0,0,0,1}     % Black
\definecolor{TUMWhite}{cmyk}{0,0,0,0}     % White

\definecolor{TUMBlue} {cmyk}{1,0.43,0,0}  % Pantone 300 C

\definecolor{TUMDarkBlue}   {cmyk}{1,0.57,0.12,0.7}      % Pantone 540 C 
\definecolor{TUMDarkerBlue} {cmyk}{1,0.54,0.04,0.19}     % Pantone 301 C
\definecolor{TUMMediumBlue} {cmyk}{0.9,0.48,0,0}         % Pantone 285 C
\definecolor{TUMLighterBlue}{cmyk}{0.65,0.19,0.01,0.04}  % Pantone 542 C
\definecolor{TUMLightBlue}  {cmyk}{0.42,0.09,0,0}        % Pantone 283 C

\definecolor{TUMDarkGray}  {cmyk}{0,0,0,0.8}  % DarkGray    80% Black
\definecolor{TUMMediumGray}{cmyk}{0,0,0,0.5}  % MediumGray  50% Black
\definecolor{TUMLightGray} {cmyk}{0,0,0,0.2}  % LightGray   20% Black

\definecolor{TUMGreen} {cmyk}{0.35,0,1,0.2}         % Pantone 383 C
\definecolor{TUMOrange}{cmyk}{0,0.65,0.95,0}        % Pantone 158 C
\definecolor{TUMIvory} {cmyk}{0.03,0.04,0.14,0.08}  % Pantone 7527 C

\definecolor{TUMBeamerYellow}    {rgb}{1.00,0.71,0.00}  % RGB 255,180,000
\definecolor{TUMBeamerOrange}    {rgb}{1.00,0.50,0.00}  % RGB 255,128,000
\definecolor{TUMBeamerRed}       {rgb}{0.90,0.20,0.09}  % RGB 229,052,024
\definecolor{TUMBeamerDarkRed}   {rgb}{0.79,0.13,0.25}  % RGB 202,033,063
\definecolor{TUMBeamerBlue}      {rgb}{0.00,0.60,1.00}  % RGB 000,153,255
\definecolor{TUMBeamerLightBlue} {rgb}{0.25,0.75,1.00}  % RGB 065,190,255
\definecolor{TUMBeamerGreen}     {rgb}{0.57,0.67,0.42}  % RGB 145,172,107
\definecolor{TUMBeamerLightGreen}{rgb}{0.71,0.79,0.51}  % RGB 181,202,130

\renewcommand{\bm}{\mathbf}

\DeclareMathOperator*{\argmax}{arg\,max}

\newcommand{\pu}{\epsilon_{\scaleto{\mathsf{U}}{3.5pt}}}

\newcommand{\npilot}{n_{\scaleto{\mathsf{P}}{3.5pt}}}

\newcommand{\pt}{\epsilon_{\scaleto{\mathsf{T}}{3.5pt}}}

\newcommand{\SCLTT}{\Lambda^{\!\scaleto{\mathsf{SCL}}{3.5pt}}}

\newcommand{\outN}{h}

\newcommand{\plist}{\mathcal{L}}
\newcommand{\olist}{\mathcal{L}_{\scaleto{\mathsf{O}}{3.5pt}}}
\newcommand{\onelist}{\mathcal{L}_{\scaleto{1}{3.5pt}}}

\newcommand{\inR}{R_{\scaleto{\mathsf{I}}{3.5pt}}}
\newcommand{\outR}{R_{\scaleto{\mathsf{O}}{3.5pt}}}

\newcommand{\ENC}{\phi}
\newcommand{\inENC}{\ENC_{\scaleto{\mathsf{I}}{3.5pt}}}
\newcommand{\outENC}{\ENC_{\scaleto{\mathsf{O}}{3.5pt}}}

\newcommand{\incode}{\mathcal{C}_{\scaleto{\mathsf{I}}{3.5pt}}}
\newcommand{\outcode}{\mathcal{C}_{\scaleto{\mathsf{O}}{3.5pt}}}
\newcommand{\outcodex}[1]{\mathcal{C}_{\scaleto{\mathsf{O}}{3.5pt}}^{\scaleto{(#1)}{5.1pt}}}
\newcommand{\code}{\mathcal{C}}

\newcommand{\fieldtwo}{\mathbb{F}_{\!2}}

\newcommand*{\QEDA}{\hfill\ensuremath{\blacksquare}}

\newcommand{\Mi}{|\setX|}

\newcommand{\safemath}[2]{\newcommand{#1}{\ensuremath{#2}\xspace}}
\def\amsbb{\use@mathgroup \M@U \symAMSb}

\safemath{\bma}{\mathbf{a}}
\safemath{\bmb}{\mathbf{b}}
\safemath{\bmc}{\mathbf{c}}
\safemath{\bmd}{\mathbf{d}}
\safemath{\bme}{\mathbf{e}}
\safemath{\bmf}{\mathbf{f}}
\safemath{\bmg}{\mathbf{g}}
\safemath{\bmh}{\mathbf{h}}
\safemath{\bmi}{\mathbf{i}}
\safemath{\bmj}{\mathbf{j}}
\safemath{\bmk}{\mathbf{k}}
\safemath{\bml}{\mathbf{l}}
\safemath{\bmm}{\mathbf{m}}
\safemath{\bmn}{\mathbf{n}}
\safemath{\bmo}{\mathbf{o}}
\safemath{\bmp}{\mathbf{p}}
\safemath{\bmq}{\mathbf{q}}
\safemath{\bmr}{\mathbf{r}}
\safemath{\bms}{\mathbf{s}}
\safemath{\bmt}{\mathbf{t}}
\safemath{\bmu}{\mathbf{u}}
\safemath{\bmv}{\mathbf{v}}
\safemath{\bmw}{\mathbf{w}}
\safemath{\bmx}{\mathbf{x}}
\safemath{\bmy}{\mathbf{y}}
\safemath{\bmz}{\mathbf{z}}
\safemath{\bmzero}{\mathbf{0}}
\safemath{\bmone}{\mathbf{1}}

\bmdefine{\biad}{a}
\bmdefine{\bibd}{b}
\bmdefine{\bicd}{c}
\bmdefine{\bidd}{d}
\bmdefine{\bied}{e}
\bmdefine{\bifd}{f}
\bmdefine{\bigd}{g}
\bmdefine{\bihd}{h}
\bmdefine{\biid}{i}
\bmdefine{\bijd}{j}
\bmdefine{\bikd}{k}
\bmdefine{\bild}{l}
\bmdefine{\bimd}{m}
\bmdefine{\bind}{n}
\bmdefine{\biod}{o}
\bmdefine{\bipd}{p}
\bmdefine{\biqd}{q}
\bmdefine{\bird}{r}
\bmdefine{\bisd}{s}
\bmdefine{\bitd}{t}
\bmdefine{\biud}{u}
\bmdefine{\bivd}{v}
\bmdefine{\biwd}{w}
\bmdefine{\bixd}{x}
\bmdefine{\biyd}{y}
\bmdefine{\bizd}{z}

\bmdefine{\bixid}{\xi}
\bmdefine{\bilambdad}{\lambda}
\bmdefine{\bimud}{\mu}
\bmdefine{\bithetad}{\theta}
\bmdefine{\biphid}{\phi}
\bmdefine{\vecalpha}{\alpha}

\safemath{\bmia}{\biad}
\safemath{\bmib}{\bibd}
\safemath{\bmic}{\bicd}
\safemath{\bmid}{\bidd}
\safemath{\bmie}{\bied}
\safemath{\bmif}{\bifd}
\safemath{\bmig}{\bigd}
\safemath{\bmih}{\bihd}
\safemath{\bmii}{\biid}
\safemath{\bmij}{\bijd}
\safemath{\bmik}{\bikd}
\safemath{\bmil}{\bild}
\safemath{\bmim}{\bimd}
\safemath{\bmin}{\bind}
\safemath{\bmio}{\biod}
\safemath{\bmip}{\bipd}
\safemath{\bmiq}{\biqd}
\safemath{\bmir}{\bird}
\safemath{\bmis}{\bisd}
\safemath{\bmit}{\bitd}
\safemath{\bmiu}{\biud}
\safemath{\bmiv}{\bivd}
\safemath{\bmiw}{\biwd}
\safemath{\bmix}{\bixd}
\safemath{\bmiy}{\biyd}
\safemath{\bmiz}{\bizd}

\safemath{\bmxi}{\bixid}
\safemath{\bmlambda}{\bilambdad}
\safemath{\bmmu}{\bimud}
\safemath{\bmtheta}{\bithetad}
\safemath{\bmphi}{\biphid}

\safemath{\bA}{\mathbf{A}}
\safemath{\bB}{\mathbf{B}}
\safemath{\bC}{\mathbf{C}}
\safemath{\bD}{\mathbf{D}}
\safemath{\bE}{\mathbf{E}}
\safemath{\bF}{\mathbf{F}}
\safemath{\bG}{\mathbf{G}}
\safemath{\bH}{\mathbf{H}}
\safemath{\bI}{\mathbf{I}}
\safemath{\bJ}{\mathbf{J}}
\safemath{\bK}{\mathbf{K}}
\safemath{\bL}{\mathbf{L}}
\safemath{\bM}{\mathbf{M}}
\safemath{\bN}{\mathbf{N}}
\safemath{\bO}{\mathbf{O}}
\safemath{\bP}{\mathbf{P}}
\safemath{\bQ}{\mathbf{Q}}
\safemath{\bR}{\mathbf{R}}
\safemath{\bS}{\mathbf{S}}
\safemath{\bT}{\mathbf{T}}
\safemath{\bU}{\mathbf{U}}
\safemath{\bV}{\mathbf{V}}
\safemath{\bW}{\mathbf{W}}
\safemath{\bX}{\mathbf{X}}
\safemath{\bY}{\mathbf{Y}}
\safemath{\bZ}{\mathbf{Z}}

\safemath{\bZero}{\mathbf{0}}
\safemath{\bOne}{\mathbf{1}}
\safemath{\bDelta}{\mathbf{\Delta}}
\safemath{\bLambda}{\mathbf{\UpLambda}}
\safemath{\bPhi}{\mathbf{\Upphi}}
\safemath{\bSigma}{\mathbf{\Upsigma}}
\safemath{\bOmega}{\mathbf{\Upomega}}
\safemath{\bTheta}{\mathbf{\Uptheta}}

\bmdefine{\biAd}{A}
\bmdefine{\biBd}{B}
\bmdefine{\biCd}{C}
\bmdefine{\biDd}{D}
\bmdefine{\biEd}{E}
\bmdefine{\biFd}{F}
\bmdefine{\biGd}{G}
\bmdefine{\biHd}{H}
\bmdefine{\biId}{I}
\bmdefine{\biJd}{J}
\bmdefine{\biKd}{K}
\bmdefine{\biLd}{L}
\bmdefine{\biMd}{M}
\bmdefine{\biOd}{O}
\bmdefine{\biPd}{P}
\bmdefine{\biQd}{Q}
\bmdefine{\biRd}{R}
\bmdefine{\biSd}{S}
\bmdefine{\biTd}{T}
\bmdefine{\biUd}{U}
\bmdefine{\biVd}{V}
\bmdefine{\biWd}{W}
\bmdefine{\biXd}{X}
\bmdefine{\biYd}{Y}
\bmdefine{\biZd}{Z}

\bmdefine{\biDelta}{\Delta}
\bmdefine{\biLambda}{\Lambda}
\bmdefine{\biPhi}{\Phi}
\bmdefine{\biSigma}{\Sigma}
\bmdefine{\biOmega}{\Omega}
\bmdefine{\biTheta}{\Theta}

\safemath{\bimA}{\biAd}
\safemath{\bimB}{\biBd}
\safemath{\bimC}{\biCd}
\safemath{\bimD}{\biDd}
\safemath{\bimE}{\biEd}
\safemath{\bimF}{\biFd}
\safemath{\bimG}{\biGd}
\safemath{\bimH}{\biHd}
\safemath{\bimI}{\biId}
\safemath{\bimJ}{\biJd}
\safemath{\bimK}{\biKd}
\safemath{\bimL}{\biLd}
\safemath{\bimM}{\biMd}
\safemath{\bimN}{\biNd}
\safemath{\bimO}{\biOd}
\safemath{\bimP}{\biPd}
\safemath{\bimQ}{\biQd}
\safemath{\bimR}{\biRd}
\safemath{\bimS}{\biSd}
\safemath{\bimT}{\biTd}
\safemath{\bimU}{\biUd}
\safemath{\bimV}{\biVd}
\safemath{\bimW}{\biWd}
\safemath{\bimX}{\biXd}
\safemath{\bimY}{\biYd}
\safemath{\bimZ}{\biZd}

\safemath{\bimDelta}{\biDelta}
\safemath{\bimLambda}{\biLambda}
\safemath{\bimPhi}{\biPhi}
\safemath{\bimSigma}{\biSigma}
\safemath{\bimOmega}{\biOmega}
\safemath{\bimTheta}{\biTheta}

\safemath{\setA}{\mathcal{A}}
\safemath{\setB}{\mathcal{B}}
\safemath{\setC}{\mathcal{C}}
\safemath{\setD}{\mathcal{D}}
\safemath{\setE}{\mathcal{E}}
\safemath{\setF}{\mathcal{F}}
\safemath{\setG}{\mathcal{G}}
\safemath{\setH}{\mathcal{H}}
\safemath{\setI}{\mathcal{I}}
\safemath{\setJ}{\mathcal{J}}
\safemath{\setK}{\mathcal{K}}
\safemath{\setL}{\mathcal{L}}
\safemath{\setM}{\mathcal{M}}
\safemath{\setN}{\mathcal{N}}
\safemath{\setO}{\mathcal{O}}
\safemath{\setP}{\mathcal{P}}
\safemath{\setQ}{\mathcal{Q}}
\safemath{\setR}{\mathcal{R}}
\safemath{\setS}{\mathcal{S}}
\safemath{\setT}{\mathcal{T}}
\safemath{\setU}{\mathcal{U}}
\safemath{\setV}{\mathcal{V}}
\safemath{\setW}{\mathcal{W}}
\safemath{\setX}{\mathcal{X}}
\safemath{\setY}{\mathcal{Y}}
\safemath{\setZ}{\mathcal{Z}}
\safemath{\emptySet}{\varnothing}

\safemath{\colA}{\mathscr{A}}
\safemath{\colB}{\mathscr{B}}
\safemath{\colC}{\mathscr{C}}
\safemath{\colD}{\mathscr{D}}
\safemath{\colE}{\mathscr{E}}
\safemath{\colF}{\mathscr{F}}
\safemath{\colG}{\mathscr{G}}
\safemath{\colH}{\mathscr{H}}
\safemath{\colI}{\mathscr{I}}
\safemath{\colJ}{\mathscr{J}}
\safemath{\colK}{\mathscr{K}}
\safemath{\colL}{\mathscr{L}}
\safemath{\colM}{\mathscr{M}}
\safemath{\colN}{\mathscr{N}}
\safemath{\colO}{\mathscr{O}}
\safemath{\colP}{\mathscr{P}}
\safemath{\colQ}{\mathscr{Q}}
\safemath{\colR}{\mathscr{R}}
\safemath{\colS}{\mathscr{S}}
\safemath{\colT}{\mathscr{T}}
\safemath{\colU}{\mathscr{U}}
\safemath{\colV}{\mathscr{V}}
\safemath{\colW}{\mathscr{W}}
\safemath{\colX}{\mathscr{X}}
\safemath{\colY}{\mathscr{Y}}
\safemath{\colZ}{\mathscr{Z}}

\safemath{\opA}{\mathbb{A}}
\safemath{\opB}{\mathbb{B}}
\safemath{\opC}{\mathbb{C}}
\safemath{\opD}{\mathbb{D}}
\safemath{\opE}{\mathbb{E}}
\safemath{\opF}{\mathbb{F}}
\safemath{\opG}{\mathbb{G}}
\safemath{\opH}{\mathbb{H}}
\safemath{\opI}{\mathbb{I}}
\safemath{\opJ}{\mathbb{J}}
\safemath{\opK}{\mathbb{K}}
\safemath{\opL}{\mathbb{L}}
\safemath{\opM}{\mathbb{M}}
\safemath{\opN}{\mathbb{N}}
\safemath{\opO}{\mathbb{O}}
\safemath{\opP}{\mathbb{P}}
\safemath{\opQ}{\mathbb{Q}}
\safemath{\opR}{\mathbb{R}}
\safemath{\opS}{\mathbb{S}}
\safemath{\opT}{\mathbb{T}}
\safemath{\opU}{\mathbb{U}}
\safemath{\opV}{\mathbb{V}}
\safemath{\opW}{\mathbb{W}}
\safemath{\opX}{\mathbb{X}}
\safemath{\opY}{\mathbb{Y}}
\safemath{\opZ}{\mathbb{Z}}
\safemath{\opZero}{\mathbb{O}}
\safemath{\identityop}{\opI}

\safemath{\veca}{\bma}
\safemath{\vecb}{\bmb}
\safemath{\vecc}{\bmc}
\safemath{\vecd}{\bmd}
\safemath{\vece}{\bme}
\safemath{\vecf}{\bmf}
\safemath{\vecg}{\bmg}
\safemath{\vech}{\bmh}
\safemath{\veci}{\bmi}
\safemath{\vecj}{\bmj}
\safemath{\veck}{\bmk}
\safemath{\vecl}{\bml}
\safemath{\vecm}{\bmm}
\safemath{\vecn}{\bmn}
\safemath{\veco}{\bmo}
\safemath{\vecp}{\bmp}
\safemath{\vecq}{\bmq}
\safemath{\vecr}{\bmr}
\safemath{\vecs}{\bms}
\safemath{\vect}{\bmt}
\safemath{\vecu}{\bmu}
\safemath{\vecv}{\bmv}
\safemath{\vecw}{\bmw}
\safemath{\vecx}{\bmx}
\safemath{\vecy}{\bmy}
\safemath{\vecz}{\bmz}

\safemath{\veczero}{\bmzero}
\safemath{\vecone}{\bmone}
\safemath{\vecxi}{\bmxi}
\safemath{\veclambda}{\bmlambda}
\safemath{\vecmu}{\bmmu}
\safemath{\vectheta}{\bmtheta}
\safemath{\vecphi}{\bmphi}

\safemath{\matA}{\bA}
\safemath{\matB}{\bB}
\safemath{\matC}{\bC}
\safemath{\matD}{\bD}
\safemath{\matE}{\bE}
\safemath{\matF}{\bF}
\safemath{\matG}{\bG}
\safemath{\matH}{\bH}
\safemath{\matI}{\bI}
\safemath{\matJ}{\bJ}
\safemath{\matK}{\bK}
\safemath{\matL}{\bL}
\safemath{\matM}{\bM}
\safemath{\matN}{\bN}
\safemath{\matO}{\bO}
\safemath{\matS}{\bS}
\safemath{\matT}{\bT}
\safemath{\matU}{\bU}
\safemath{\matW}{\bW}
\safemath{\matX}{\bX}
\safemath{\matzero}{\bmzero}

\safemath{\matDelta}{\bDelta}
\safemath{\matLambda}{\bLambda}
\safemath{\matPhi}{\bPhi}
\safemath{\matSigma}{\bSigma}
\safemath{\matOmega}{\bOmega}
\safemath{\matTheta}{\bTheta}

\safemath{\matidentity}{\matI}
\safemath{\matone}{\matO}

\safemath{\rnda}{a}
\safemath{\rndb}{b}
\safemath{\rndc}{c}
\safemath{\rndd}{d}
\safemath{\rnde}{e}
\safemath{\rndf}{f}
\safemath{\rndg}{g}
\safemath{\rndh}{h}
\safemath{\rndi}{i}
\safemath{\rndj}{j}
\safemath{\rndk}{k}
\safemath{\rndl}{l}
\safemath{\rndm}{m}
\safemath{\rndn}{n}
\safemath{\rndo}{o}
\safemath{\rndp}{p}
\safemath{\rndq}{q}
\safemath{\rndr}{r}
\safemath{\rnds}{s}
\safemath{\rndt}{t}
\safemath{\rndu}{u}
\safemath{\rndv}{v}
\safemath{\rndw}{w}
\safemath{\rndx}{x}
\safemath{\rndy}{y}
\safemath{\rndz}{z}

\safemath{\rveca}{\bma}
\safemath{\rvecb}{\bmb}
\safemath{\rvecc}{\bmc}
\safemath{\rvecd}{\bmd}
\safemath{\rvece}{\bme}
\safemath{\rvecf}{\bmf}
\safemath{\rvecg}{\bmg}
\safemath{\rvech}{\bmh}
\safemath{\rveci}{\bmi}
\safemath{\rvecj}{\bmj}
\safemath{\rveck}{\bmk}
\safemath{\rvecl}{\bml}
\safemath{\rvecm}{\bmm}
\safemath{\rvecn}{\bmn}
\safemath{\rveco}{\bmo}
\safemath{\rvecp}{\bmp}
\safemath{\rvecq}{\bmq}
\safemath{\rvecr}{\bmr}
\safemath{\rvecs}{\bms}
\safemath{\rvect}{\bmt}
\safemath{\rvecu}{\bmu}
\safemath{\rvecv}{\bmv}
\safemath{\rvecw}{\bmw}
\safemath{\rvecx}{\bmx}
\safemath{\rvecy}{\bmy}
\safemath{\rvecz}{\bmz}

\safemath{\rvecxi}{\bmxi}
\safemath{\rveclambda}{\bmlambda}
\safemath{\rvecmu}{\bmmu}
\safemath{\rvectheta}{\bmtheta}
\safemath{\rvecphi}{\bmphi}
\safemath{\rmatDelta}{\bimDelta}
\safemath{\rmatLambda}{\bimLambda}
\safemath{\rmatPhi}{\bimPhi}
\safemath{\rmatSigma}{\bimSigma}
\safemath{\rmatOmega}{\bimOmega}
\safemath{\rmatTheta}{\bimTheta}

\newcommand{\rate}{R}

\newcommand{\infden}{\imath_{s}}

\newcommand{\lefto}{\mathopen{}\left}
\newcommand{\ltrp}[1]{\ensuremath{\mathopen{}\left(#1\right)}}
\newcommand{\ltrcurley}[1]{\ensuremath{\mathopen{}\left\{#1\right\}}}
\newcommand{\ltrsqr}[1]{\ensuremath{\mathopen{}\left[#1\right]}}
\newcommand{\supp}[1]{\ensuremath{^{\subtext{#1}}}} % for text mode subscripts

\newenvironment{textbmatrix}{\setlength{\arraycolsep}{2.5pt}%
	\big[\begin{matrix}}{\end{matrix}\big]%
	\raisebox{0.08ex}{\vphantom{M}}}

\def\be{\begin{equation}}
		\def\ee{\end{equation}}
\def\een{\nonumber \end{equation}}
\def\mat{\begin{bmatrix}}
		\def\emat{\end{bmatrix}}
\def\btm{\begin{textbmatrix}}
		\def\etm{\end{textbmatrix}}

\def\ba#1\ea{\begin{align}#1\end{align}}
\def\bas#1\eas{\begin{align*}#1\end{align*}}
\def\bs#1\es{\begin{split}#1\end{split}}
\def\bg#1\eg{\begin{gather}#1\end{gather}}
\def\bi#1\ei{\begin{itemize}#1\end{itemize}}

\newcommand{\subtext}[1]{\text{\fontfamily{cmr}\fontshape{n}\fontseries{m}\selectfont{}#1}}
{
\newcommand{\sub}[1]{\ensuremath{_{\subtext{#1}}}} % for text mode subscripts

\definecolor{mpurple}{rgb}{0.58, 0.44, 0.86}

\DeclareMathOperator{\Prob}{\opP}			% probability of an event
\DeclareMathOperator{\Exop}{\opE}			% expectation operator
\newcommand{\Ex}[2]{\ensuremath{\Exop_{#1}\lefto[#2\right]}} 	% 
\newcommand{\abs}[1]{\left\lvert#1\right\rvert}		% absolute value
\newcommand{\Union}{\bigcup}
\newcommand{\ind}[1]{\mathbbm{1}(#1)}
\newcommand{\vecnorm}[1]{\lVert#1\rVert}		% vector norm
\safemath{\dirac}{\delta}					% Dirac delta
\safemath{\krond}{\dirac}					% Kronecker delta
\safemath{\upto}{\uparrow}
\safemath{\downto}{\downarrow}
\safemath{\iu}{j}							% imaginary unit
\safemath{\ev}{\lambda}						% eigenvalue
\safemath{\hilseqspace}{l^{2}}				% Hilbert sequence space
\newcommand{\banachfunspace}[1]{\setL^{#1}}	% Banach function space
\safemath{\hilfunspace}{\banachfunspace{2}}	% Hilbert function space
\newcommand{\ceil}[1]{\lceil #1 \rceil}  	% Ceil operator

\safemath{\snr}{\rho} 				% signal to noise ratio
\safemath{\No}{N_0}							% noise spectral density
\safemath{\Es}{E_s}							% energy per symbol
\safemath{\Eb}{E_b}							% energy per bit
\safemath{\EbNo}{\frac{\Eb}{\No}}
\safemath{\EsNo}{\frac{\Es}{\No}}

\DeclareMathOperator{\CHop}{\ensuremath{\opH}} % channel operator
\safemath{\tvir}{\rndh_{\CHop}}				% time-varying impulse response
\safemath{\tvtf}{\rndl_{\CHop}}				% 	-''- transfer function
\safemath{\spf}{\rnds_{\CHop}}				% spreading function
\safemath{\bff}{H_{\CHop}}					% bi-freuqency function

\safemath{\ircf}{r_{h}}						% impulse response correlation fn.
\safemath{\tftvcf}{r_{s}}					% scattering function
\safemath{\tfcf}{r_{l}}						% time-frequency correlation fn.
\safemath{\bfcf}{r_{H}}						% bi-frequency correlation fn.

\safemath{\tcorr}{c_h}						% time-correlation function
\safemath{\scf}{c_{s}}						% spreading function
\safemath{\tfcorr}{c_{l}}					% transfer-function correlation
\safemath{\fcorr}{c_{H}}						% frequency-correlation function

\safemath{\mi}{I}							% mutual information
\safemath{\capacity}{C}						% capacity
\safemath{\difent}{h}						% differential entropy
\newcommand{\iid}{i.i.d.\@\xspace}

\safemath{\normal}{\mathcal{N}}			% normal distribution
\safemath{\jpg}{\mathcal{CN}}			% jointly proper Gaussian
\safemath{\mchain}{\leftrightarrow}		% Markov chain
\newcommand{\given}{\,\vert\,}				% conditioning
\newcommand{\ggiven}{\;\middle|\;}             % Conditioning for direct use

\safemath{\wpone}{\text{w.p.1}}		% with probability 1

\safemath{\dB}{\,\mathrm{dB}}
\safemath{\dBm}{\,\mathrm{dBm}}
\safemath{\Hz}{\,\mathrm{Hz}}
\safemath{\kHz}{\,\mathrm{kHz}}
\safemath{\MHz}{\,\mathrm{MHz}}
\safemath{\GHz}{\,\mathrm{GHz}}
\safemath{\s}{\,\mathrm{s}}
\safemath{\ms}{\,\mathrm{ms}}
\safemath{\mus}{\,\mathrm{\text{\textmu}s}}
\safemath{\ps}{\,\mathrm{ps}}
\safemath{\meter}{\,\mathrm{m}}
\safemath{\mm}{\,\mathrm{mm}}
\safemath{\cm}{\,\mathrm{cm}}
\safemath{\m}{\,\mathrm{m}}
\safemath{\W}{\,\mathrm{W}}
\safemath{\mW}{\, \mathrm{mW}}
\safemath{\J}{\,\mathrm{J}}
\safemath{\K}{\,\mathrm{K}}
\safemath{\bit}{\,\mathrm{bit}}
\safemath{\nat}{\,\mathrm{nat}}
\safemath{\kg}{\,\mathrm{kg}}

\safemath{\define}{\triangleq}			% definition

\safemath{\equivalent}{\sim}
\safemath{\distas}{\sim}					% distributed according to
\safemath{\sdiff}{\Delta}				% symmetric set difference

\safemath{\reals}{\amsbb{R}}
\safemath{\positivereals}{\reals_{+}}
\safemath{\integers}{\amsbb{Z}}
\safemath{\posint}{\integers_{+}}
\safemath{\naturals}{\mathbb{N}}
\safemath{\posnaturals}{\naturals_{+}}
\safemath{\complexset}{\amsbb{C}}
\safemath{\rationals}{\amsbb{Q}}
\safemath{\bifield}{\amsbb{F}_2}

\newcommand*{\fancyrefapplabelprefix}{app}		% Appendix
\newcommand*{\fancyrefthmlabelprefix}{thm}		% Theorem
\newcommand*{\fancyreflemlabelprefix}{lem}		% Lemma
\newcommand*{\fancyrefcorlabelprefix}{cor}		% Corolary
\newcommand*{\fancyrefdeflabelprefix}{def}		% Definition
\newcommand*{\fancyrefproplabelprefix}{prop}		% Property
\newcommand*{\fancyrefexelabelprefix}{exe}		% Exercise
\newcommand*{\fancyrefpropolabelprefix}{propo}		% Proposition
\frefformat{vario}{\fancyrefseclabelprefix}{Section~#1}
\frefformat{vario}{\fancyrefthmlabelprefix}{Theorem~#1}
\frefformat{vario}{\fancyreflemlabelprefix}{Lemma~#1}
\frefformat{vario}{\fancyrefcorlabelprefix}{Corollary~#1}
\frefformat{vario}{\fancyrefdeflabelprefix}{Definition~#1}
\frefformat{vario}{\fancyreffiglabelprefix}{Fig.~#1}
\frefformat{vario}{\fancyrefapplabelprefix}{Appendix~#1}
\frefformat{vario}{\fancyrefeqlabelprefix}{(#1)}
\frefformat{vario}{\fancyrefproplabelprefix}{Property~#1}
\frefformat{vario}{\fancyrefexelabelprefix}{Exercise~#1}
\frefformat{vario}{\fancyrefpropolabelprefix}{Propo~#1}

\newtheorem{thm}{Theorem}

\newtheorem{rem}{Remark}

\newtheorem{dfn}{Definition}

\colorlet{lgray}{gray!30}

\makeatletter
\newcommand{\vast}{\bBigg@{3}}
\newcommand{\Vast}{\bBigg@{4}}
\makeatother

\IEEEoverridecommandlockouts

\begin{document}

%%%%%%%%%%%%%%%%%%%%%%%%%%%%%%%%%%%%%%%%%%%%%%%%%%%%%%%%%%%%%%%%%%%%%%%%%%%%%%%%%%%%%%%%%%%%%%%%%%%%%%%%%%%%%%%%%%%%%%%%%%%%%%%%%%
\title{Undetected Error Probability in the Short Blocklength Regime: Approaching Finite-Blocklength Bounds with Polar Codes}

\author{Alexander~Sauter,~\IEEEmembership{Graduate Student Member,~IEEE}, A.~Oguz~Kislal, Giuseppe~Durisi,~\IEEEmembership{Senior Member,~IEEE},\\ Gianluigi~Liva,~\IEEEmembership{Senior Member,~IEEE},
Bal\'azs~Matuz,~\IEEEmembership{Senior Member,~IEEE}, and Erik~G.~Str\"om,~\IEEEmembership{Fellow,~IEEE}\\

\thanks{
	This paper was presented in part at the 57th Annual Conference on Information Sciences and Systems (CISS)~\cite{Sauter2023}.}
\thanks{
Alexander Sauter is with the Institute for Communications Engineering, School of Computation, Information, and Technology, Technical University of Munich, Arcisstrasse 21, D-80333 Munich, Germany, and with the German Aerospace Center, M\"unchener Strasse 20, D-82234 Wessling, Germany. Gianluigi~Liva is with the German Aerospace Center, M\"unchener Strasse 20, D-82234 Wessling, Germany (e-mail: {\tt \{alexander.sauter,gianluigi.liva,\}@dlr.de}). A. Oguz Kislal, Giuseppe Durisi, and Erik. G. Str\"om are with the Department of Electrical Engineering, Chalmers University of Technology, Gothenburg 41296, Sweden (e-mail: {\tt \{kislal,durisi,erik.strom\}@chalmers.se}). Bal\'{a}zs Matuz is with the Huawei Munich Research Center, 80992 Munich, Germany (e-mail: {\tt balazs.matuz@huawei.com}).}
\thanks{
	Alexander Sauter and Gianluigi Liva acknowledge the financial support by the Federal Ministry of Education and Research of Germany in the program of ``Souver\"an. Digital. Vernetzt.'', by the joint project 6G-RIC, project identification number: 16KISK022, and by the DLR-funded project ``DIAL''.
}
\thanks{This paper was supported in part by the Swedish Research Council under grants 2021-04970 and 2022-04471.}
}

\maketitle

\IEEEoverridecommandlockouts

\begin{abstract}
	We analyze the trade-off between the undetected error probability (i.e., the probability that the channel decoder outputs an erroneous message without detecting the error) and the total error probability in the short blocklength regime. We address the problem by developing two new finite blocklength achievability bounds, which we use to benchmark the performance of two coding schemes based on polar codes with outer cyclic redundancy check (CRC) codes---also referred to as  CRC-aided (CA) polar codes.
	The first bound is obtained by considering an outer detection code, whereas the second bound relies on a threshold test applied to the generalized information density. Similarly, in the first CA polar code scheme, we reserve a fraction of the outer CRC parity bits for error detection, whereas in the second scheme, we apply a threshold test (specifically, Forney's optimal rule) to the output of the successive cancellation list decoder.
	Numerical simulations performed on the binary-input AWGN channel reveal that, in the short-blocklength regime, the threshold-based approach is superior to the CRC-based approach, both in terms of bounds and performance of CA polar code schemes.
	We also consider the case of decoding with noisy channel-state information, which leads to a mismatched decoding setting.
	Our results illustrate that, differently from the previous case, in this scenario, the CRC-based approach outperforms the threshold-based approach, which is more sensitive to the mismatch.
\end{abstract}


%%%%%%%%%%%%%%%%%%%%%%%%%%%%%%%%%%%%%%%%%%%%%%%%%%%%%%%%%%%%%%%%%%%%%%%%%%%%%%%%%%%%%%%%%%%%%%%%%%%%%%%%%%%%%%%%%%%%%%%%%%%%%%%%%%

\begin{acronym}

	\acro{ASCL}{augmented successive cancellation list}
	\acro{AWGN}{additive white Gaussian noise}

	\acro{BEC}{binary erasure channel}
	\acro{biAWGN}{binary-input additive white Gaussian noise}
	\acro{BP}{belief propagation}
	\acro{BPSK}{binary phase shift keying}
	\acro{BSC}{binary symmetric channel}
	\acro{BPCU}{bit per channel use}
	\acro{CCC}{constant composition code}
	\acro{CN}{check node}
	\acro{CRC}{cyclic redundancy check}
	\acro{CSI}{channel state information}

	\acro{DE}{density evolution}
	\acro{DM}{distribution matcher}
	\acro{DMC}{discrete memoryless channel}

	\acro{eMBB}{enhanced mobile broadband}
	\acro{EXIT}{extrinsic information transfer}

	\acro{GA}{Gaussian approximation}

	\acro{LEO}{low earth orbit}
	\acro{LDPC}{low-density parity-check}

	\acro{ML}{maximum likelihood}
	\acro{mMTC}{massive machine-type communication}
	\acro{MN}{MacKay-Neal}

	\acro{NS}{non-systematic}

	\acro{OOK}{on-off keying}

	\acro{PAT}{pilot-assisted transmission}
	\acro{PEXIT}{protograph extrinsic information transfer}
	\acro{PEP}{pairwise error probability}
	\acro{p.m.f.}{probability mass function}
	\acro{p.d.f.}{probability density function}

	\acro{QPSK}{quadrature phase-shift keying}

	\acro{RA}{repeat-accumulate}
	\acro{RCB}{random coding bound}
	\acro{RCU}{random coding union}
	\acro{r.v.}{random variable}

	\acro{SC}{successive cancellation}
	\acro{SCL}{successive cancellation list}
	\acro{SNR}{signal-to-noise ratio}

	\acro{TEP}{total error probability}

	\acro{UEP}{undetected error probability}
	\acro{URLLC}{ultra-reliable low-latency communications}

	\acro{VN}{variable node}

\end{acronym}

%%%%%%%%%%%%%%%%%%%%%%%%%%%%%%%%%%%%%%%%%%%%%%%%%%%%%%%%%%%%%%%%%%%%%%%%%%%%%%%%%%%%%%%%%%%%%%%%%%%%%%%%%%%%%%%%%%%%%%%%%%%%%%%%%%
\section{Introduction}
\IEEEPARstart{T}{he} design of efficient short error correcting codes is subject of
renewed interest due to emerging applications envisaged in $5$G and
beyond~\cite{Durisi2016,GLP2021}. Consider as an example the $3$GPP $5$G NR standard. In
the case of \ac{eMBB} links, small data units play an essential role in the control
channel. In the \ac{mMTC} setting, a large number of devices transmit short packets in a
sporadic and uncoordinated manner. In \ac{URLLC}, delay constraints require the use of
short error correcting codes. Furthermore, strict reliability requirements call for low
post-decoding error rates. The \ac{URLLC} scenario is especially relevant in the context
of intelligent mobility, industry automation, cyber-physical systems, and wireless
telecommand systems (see, e.g., \cite{Liva2011:CCSDS_IEEEPROC}, \cite[Chapter
	4]{CCSDS14}). 
    
    The reliability requirements of \ac{URLLC} systems are typically expressed in
terms of block error probability at the decoder output. However, this is not
sufficient in certain mission-critical applications, for which it is essential to
distinguish between two types of decoding errors: \emph{detected} and \emph{undetected}
errors.
An error is detected if the channel decoder signals to the upper
layers a decoding failure (\emph{erasure}).
An error is undetected if
the decoder outputs an erroneous message, which is forwarded to the upper layers.
In the absence of additional error detection capabilities in the upper layers---provided by, e.g., additional
\ac{CRC} codes---undetected errors can be particularly
harmful \cite{Devassy2019}. It is, hence, imperative to assess the performance of
an error correction algorithm in terms of \emph{both} the \ac{TEP} and the \ac{UEP}.

The design of channel codes capable of providing low \ac{UEP} is challenging
	at short blocklengths.
When the blocklength is large, error detection
	capabilities can be provided by appending to the data packet a sufficiently long \ac{CRC} code, which is used
for error detection after decoding. The key observation is that, for long packets, the
addition of the \ac{CRC} code parity bits causes only a negligible rate loss, and,
	hence, a small \ac{TEP} penalty.
On the contrary, for short
packets, the use of a \ac{CRC} code as error detection
mechanism\footnote{\ac{CRC} codes are sometimes used, in concatenation with short polar
	codes, to improve the error correction performance of the inner polar code under
	successive cancellation list decoding~\cite{Tal2013}. This setting
	should not be confused with the one where the outer \ac{CRC}
	is used purely for error detection.} can
result in an unacceptable rate loss and, hence, a significant \ac{TEP} penalty.
A more appealing
solution is the use of an \emph{incomplete}\footnote{A complete decoder is a
	decoder returning always a valid codeword. In contrast, an incomplete
	decoder returns a valid codeword or an erasure, i.e., a decoding failure flag. Any blockwise maximum likelihood decoder or the
	successive cancellation decoder of polar codes are examples of complete decoders.
	Decoders based on the belief propagation algorithm and bounded-distance decoders are incomplete decoders. See, e.g., \cite[Ch. 1]{blahut2003algebraic}.}
decoder, capable of detecting decoding errors with a sufficiently high probability.

An optimum incomplete decoding rule was introduced and analyzed using
	information-theoretic tools by Forney in~\cite{For68}. The approach of \cite{For68} can be interpreted as the application of
(complete) \ac{ML} decoding, followed by a post-decoding threshold test. The test is used
to either accept or discard the \ac{ML} decoder decision. More specifically, Forney
demonstrated that the decoder that optimally trades  between \ac{TEP} and \ac{UEP}
operates as follows: it outputs the message whose likelihood is at least
$2^{nT}$ times larger than the sum of the likelihood of all other messages. If
no message satisfies this condition, it declares an erasure. Here, $n$ denotes the
blocklength, and $T$ is a suitably chosen threshold. Forney also performed an
error-exponent analysis of this decoding rule, demonstrating that both \ac{TEP} and
\ac{UEP} decay exponentially fast with the blocklength for \ac{DMC}. Forney's decoding
metric can be efficiently evaluated for certain code classes (e.g., for terminated
convolutional codes \cite{Raghavan1998,Hof2009:ISTCA}), or it can be well approximated for
other code classes (e.g., for codes based on compact tail-biting trellises
\cite{Williamson2014:ROVA,Wesel2022:ROVA}).
Suboptimal tests are proposed and analyzed in \cite{Dolinar2008:Angle,Dolinar2008:Bounds}.
Heuristic threshold tests are introduced in \cite{Jang2021:UDER_Polar} for CA polar codes
\cite{Sto02,Ari09} under \ac{SCL} decoding \cite{TV15}. Error detection for polar codes under \ac{SCL} via typicality check or by using a random linear code as outer error detection code was analyzed in \cite{Pedarsani11}.

Forney's error-exponent analysis has been extended in various directions in the literature. In
\cite{Merhav2008}, the author provides error exponent bounds that are at least as tight as Forney's error exponent bounds and are simpler to evaluate in certain cases.
This analysis was further extended in~\cite{somekh-baruch11-10a}, where the exact random coding
	exponent is determined for the \iid codebook ensemble under certain channel-symmetry conditions.
In
\cite{Barg2004}, the author derives \ac{TEP} and \ac{UEP} bounds for spherical codes over
the \ac{AWGN} channel under a suboptimal decoding rule. The authors of \cite{Hof2010:IT}
provide bounds for structured codes.
Specifically, using the distance distribution of a linear block
code ensemble, they improve Forney's bounds for some linear block code ensembles. In
\cite{TelatarPHD,Telatar1994,Moulin2008}, Forney's result is generalized for the
case in which constant composition codes are used.
All these results, however, rely on
Gallager's error exponent framework, which typically yields loose bounds for short
blocklengths and error probabilities of interest for \ac{URLLC} applications~\cite{Polyanskiy2010,Lancho2020}.

In~\cite{Hayashi2015}, erasure decoding is analyzed in the moderate deviation regime,
i.e., the regime in which the code rate tends to capacity at a rate slower than
$n^{-1/2}$ and the error probability decays with $n$.\footnote{As a comparison, in the error-exponent regime, the rate is fixed and the error probability decays exponentially with $n$. Another regime of interest is the second-order asymptotic regime, where the rate converges to capacity at a rate of $n^{-1/2}$ and the error probability is fixed.}
Notably, this analysis employs a suboptimal decoder that involves thresholding
the information density rather than Forney's optimal metric, which is not
	analytically tractable in this regime.
In~\cite{Vincent2014},
the second-order term in the asymptotic expansion of the maximum coding rate, for a given
fixed error probability, in the
asymptotic limit of large blocklength is evaluated for erasure decoding. Similar to~\cite{Hayashi2015}, the achievability part of the proof relies on a suboptimal threshold
decoder. The analyses in \cite{Hayashi2015} and \cite{Vincent2014} do not aim to obtain
numerically computable bounds for short blocklengths.
Hence, they do not directly provide ways to benchmark the performance of actual
	coding schemes in the short-packet regime.
As we shall see,
employing a suboptimal decoder that involves thresholding the (generalized)
information density will prove to be an effective approach to obtain numerically
	computable bounds.

\smallskip

\paragraph*{Contributions}
In this work, we study the \ac{TEP} and the \ac{UEP} for short polar codes, concatenated
with outer \ac{CRC} codes, which we shall refer to as CRC-aided (CA) polar codes, in
the following. We focus on this code class because of its excellent performance in
the short block length regime with low-complexity \ac{SCL} decoding
\cite{Liva2019:Survey}. We introduce two error detection methods. A first method
relies on ``splitting'' the parity bits of the \ac{CRC} code: a portion of the bits
	is used to prune the \ac{SCL} decoder list, whereas the remaining parity bits are used for
	error detection. The second approach is based on the optimal threshold test of~\cite{For68},
	adapted to \ac{SCL} decoding. The \ac{TEP} and \ac{UEP} performance under the two
approaches are
analyzed over the \ac{biAWGN} channel, and over the block-memoryless phase-noise
channel \cite{Peleg98} with imperfect \ac{CSI} at the receiver.
To benchmark the performance of these short polar codes, we devise two novel
	information-theoretic achievability bounds.
Specifically, in agreement with the first detection method, the first bound is
	obtained by using an outer decoder for error detection.
Furthermore, in agreement with the second detection method, the second bound relies on a threshold decoder (although, following~\cite{Hayashi2015,Vincent2014}, a suboptimal decoder is used for analytical tractability).
Unlike Forney's bound,
both bounds are based on the \ac{RCU} bound \cite[Thm.~16]{Polyanskiy2010}. The \ac{RCU} bound
requires the evaluation of a certain tail probability, which is not known in closed
form and must be evaluated numerically. This is extremely time-consuming due to the
low error probabilities of interest in \ac{URLLC}. To tackle this issue, we present a
saddlepoint approximation, which generalizes the one reported in
\cite{font-segura18-03a}. Numerical experiments lead to the following observations:
\begin{itemize}
	\item The two novel achievability bounds outperform Forney's bound for the
	      \ac{biAWGN} channel in the short blocklength regime and for the \ac{TEP} and
		      \ac{UEP} considered in the paper.
		      Furthermore, the bound obtained through
		      the suboptimal threshold decoder outperforms the one based on an outer code.
		      However, the gap between the two bounds
	      becomes less pronounced as the blocklength increases. The simulation results for CA
	      polar codes over the \ac{biAWGN} channel confirm these insights: the threshold test
	      that approximates the decision metric of \cite{For68} by using the \ac{SCL} decoder
	      output outperforms the detection method based on an outer \ac{CRC}
	      for short blocklengths. Also in this case, the gain vanishes as the blocklength
	      grows.
	\item Over the block-memoryless phase-noise channel with imperfect \ac{CSI}, the gap
	      between the two achievability bounds is drastically reduced. Simulation results for
	      CA polar codes follow the trend displayed by the bounds. This suggests that,
	      in the presence of inaccurate \ac{CSI}, error detection by means of an outer CRC
		      code is more robust and, hence, preferable.
	      On the contrary, threshold-based techniques
	      that rely on a mismatched likelihood suffer from a significant performance loss
	      for a wide range of blocklengths.
\end{itemize}

\smallskip

The remainder of the manuscript is organized as follows. In
Section~\ref{sec:preliminaries}, we introduce the notation
and the system model. In
Section~\ref{sec:bounds},
we review Forney's optimum erasure decoder, as well as its error exponent analysis
\cite{For68}. Then, two new achievability bounds are presented, and their numerical
evaluation is discussed. Error detection strategies for CA polar codes are described in
Section~\ref{sec:error_det}. Numerical
results are provided in Section~\ref{sec:results}
and conclusions follow in
Section~\ref{sec:conclusions}.

%%%%%%%%%%%%%%%%%%%%%%%%%%%%%%%%%%%%%%%%%%%%%%%%%%%%%%%%%%%%%%%%%%%%%%%%%%%%%%%%%%%%%%%%%%%%%%%%%%%%%%%%%%%%%%%%%%%%%%%%%%%%%%%%%%
\section{Preliminaries}\label{sec:preliminaries}

\subsection{Notation}\label{sec:notation}
We denote vectors by small bold letters, e.g., $\bm x$, and sets by capital calligraphic
letters, e.g., $\mathcal X$. The cardinality of a set $\setX$ is denoted by $|\setX|$. We use $\fieldtwo$ for the binary finite field with elements
$\{0,1\}$ and $\mathbb{N}_0$ to denote the set of natural numbers including $0$. We write
$\log(\cdot)$  to denote the natural logarithm and $\log_2(\cdot)$ to denote the base-$2$
logarithm. Moreover, $\vecnorm{\cdot}$ stands for the $\ell_2$-norm, $\Prob[\cdot]$
for the probability of an event, $\Exop[\cdot]$ for the expectation operator, $Q(\cdot)$
for the Gaussian $Q$-function, and $\ind{\cdot}$ for the indicator function. Finally, for
two functions $f(n)$ and $g(n)$, the notation $f(n) = o(g(n))$ means that
$\lim_{n\to\infty} f(n)/g(n)=0$.

\subsection{System Model}
\label{sec:SystemModel}
We consider an arbitrary discrete-time communication channel that maps input symbols from
the set $\setX$  into output symbols from the set $\setY$. Specifically, let $\vecx =
	[x_1,\ldots,x_{n}] \in \setX^{n}$ and $\rvecy = [\rndy_{1}, \ldots, \rndy_{n}] \in
	\setY^{n}$ be vectors containing $n$ channel inputs and their corresponding outputs.
The channel is
defined by its transition probability\footnote{To keep the notation simple, we have chosen
	to specify the transition probability in terms of the conditional probability mass
	function $P_{\rvecy|\rvecx}(\rvecy|\rvecx)$, which requires the set $\setY$ to have
	finite cardinality. However, our analysis is general and can be applied also to channels
	with continuous input and output. In this case, $P_{\rvecy|\rvecx}(\rvecy|\rvecx)$ should
	be replaced by the conditional probability density function
	$p_{\rvecy|\rvecx}(\rvecy|\rvecx)$.} $P_{\rvecy|\rvecx}(\rvecy|\rvecx)$.

We next define the notion of a channel code.
Similar to, e.g.,~\cite{Polyanskiy2011}, it will turn out convenient to focus
	on the class of \emph{randomized} coding schemes, i.e., coding schemes in which a common source
	of randomness (which we denote by $u$) is available at both transmitter and receiver.
	This randomness is used to initialize the encoder and decoder.
\begin{dfn}\label{def:code}
	An $(n,k,\pt,\pu)$-code for the channel $P_{\rvecy|\rvecx}$ consists of
	\begin{itemize}
		\item A discrete random variable $\rndu$ with distribution $P_{\rndu}$ defined on a set $\setU$ with $\abs{\setU} \leq 2$ that is revealed to both the transmitter and the receiver before the start of transmission. This allows the transmitter and the receiver to time-share between up to two deterministic codes.

		\item An encoder $\ENC: \setU \times \{1,\ldots,2^{k}\} \rightarrow \setX^{n}$ that
		      maps a message $\rndw$, which we assume to be uniformly distributed on
		      $\{1,\ldots,2^{k}\}$, to a codeword in the set $\setC_u \subset \setX^{n}$.

		\item An erasure decoder $g:\setU \times \setY^{n} \rightarrow\{0,1,\ldots,2^{k}\}$
		      that maps the received vector to one of the messages in $\{1,\ldots, 2^{k}\}$, or
		      declares an erasure, which we indicate by the extra symbol $0$. Let, for a given
		      $u$,
		      $\setD_{\rndu,\hat{\rndw}} = g^{-1}(\rndu,\hat{\rndw}) \subset \setY^{n}$ denote the
		      decoding region associated to each decoder output $\hat{\rndw} \in \{0,1,\ldots, 2^{k}\}$
		      and assume that these decoding regions form a partition of $\setY^{n}$ for each fixed $u$.
		      We require that the \ac{TEP}  and \ac{UEP} do not
		      exceed $\pt$ and $\pu$, respectively.
		      Mathematically,
		      \begin{equation}\label{eq:TEP}
			      \frac{1}{2^k} \sum_{m=1}^{2^k} \sum_{\substack{m'=0 \\ m'\neq m}}^{2^k} \Prob\ltrsqr{\rvecy \in\setD_{\rndu,m'}| \rndw = m} \leq \pt
		      \end{equation}
		      \begin{equation}\label{eq:UEP}
			      \frac{1}{2^k} \sum_{m=1}^{2^k} \sum_{\substack{m'=1 \\ m'\neq m}}^{2^k} \Prob\ltrsqr{\rvecy \in\setD_{\rndu,m'}| \rndw = m} \leq \pu
		      \end{equation}
              where in \eqref{eq:TEP} and \eqref{eq:UEP} the probabilities are computed over the pair $(u,\vecy)$.
	\end{itemize}
\end{dfn}

%%%%%%%%%%%%%%%%%%%%%%%%%%%%%%%%%%%%%%%%%%%%%%%%%%%%%%%%%%%%%%%%%%%%%%%%%%%%%%%%%%%%%%%%%%%%%%%%%%%%%%%%%%%%%%%%%%%%%%%%%%%%%%%%%%
\section{Non-asymptotic Achievability Bounds on \ac{TEP} and \ac{UEP}}\label{sec:bounds}

In this section, we will first review Forney's optimal erasure decoder. Then, we will
introduce two novel finite-blocklength bounds, which are tailored to the short-blocklength
regime.
These two bounds rely on error detection strategies similar to the ones we will consider
for CA polar codes in Section~\ref{sec:error_det}.

\subsection{Optimum Erasure Decoder}
Forney, in his seminal paper \cite{For68}, showed that the optimal erasure
decoder,\footnote{The decoder is optimal in the sense that no other decoder can achieve
	simultaneously a lower $\pt$ and a lower $\pu$.} has the following structure: the decoder
outputs the message corresponding to the codeword $\rvecx$ that satisfies
\begin{equation}
	\label{eq:forneytest}
	\Lambda(\vecx,\vecy) > 2^{n T}
\end{equation}
where $T \geq 0$ is a parameter\footnote{When $T<0$, the decoding regions are not disjoint and multiple codewords satisfy~\eqref{eq:forneytest}.
	In this case, the decoder puts out a list of messages corresponding to all codewords satisfying~\eqref{eq:forneytest}. In agreement with Definition~\ref{def:code}, we will focus on the case $T \geq 0$.} that controls the tradeoff between $\pu$ and $\pt$, and
\begin{equation}
	\label{eq:forneytt}
	\Lambda(\vecx,\vecy) = \frac{P_{\rvecy|\rvecx}\left(\bm y |\bm x \right)}{\sum_{\bm x' \in \code\setminus  \{\vecx\}} P_{\rvecy|\rvecx}\left(\bm y |\bm x' \right) }.
\end{equation}
If no codeword in the codebook satisfies \eqref{eq:forneytest}, the decoder declares an erasure.
We note that the channel law $P_{\rvecy|\rvecx}(\rvecy|\rvecx)$ needs to be known at the
receiver to evaluate \eqref{eq:forneytt} and that the decoder puts out the \ac{ML}
decision when~\eqref{eq:forneytest} is satisfied.

A characterization of the achievable  $(\pt,\pu)$ pairs under this decoder over a \ac{DMC}
was provided in \cite{For68} through a random coding error exponent analysis (see,
e.g.,~\cite{Merhav2008,somekh-baruch11-10a} for more recent extensions). In particular,
it was shown that for an arbitrary input distribution $P_{x}$, there exists a
deterministic block code of length $n$ and rate $R = k/n$ that simultaneously satisfies
$\pt\leq \pt\supp{F}$ and $\pu\leq \pu\supp{F}$, where\footnote{Randomization via
	time-sharing is not needed to achieve~\eqref{eq:RCB_Pe} and~\eqref{eq:RCB_Pu}.}
\begin{align}
	\pt\supp{F}=2^{-n E_1(R,T,P_{x})} \label{eq:RCB_Pe} \\
	\pu\supp{F}=2^{-n E_2(R,T,P_{x})\label{eq:RCB_Pu}}
\end{align}
and where
\begin{align}
	E_1(R,T,P_{x})    & =\max_{0\leq s\leq \rho \leq 1} \left[E_0(s,\rho,P_{x})-\rho R-sT\right]                           \\[2mm]
	E_2(R,T,P_{x})    & =E_1(R,T,P_{x})+T                                                                                  \\
	\intertext{with}
	E_0(s,\rho,P_{x}) & = - \log_2 \sum_{y \in \setY} \left(\sum_{x \in \setX} P_{x}(x) P_{y|x}(y|x)^{1-s}\right)\nonumber \\
	                  & \quad \times \left(\sum_{x' \in \setX} P_{x}(x') P_{y|x}(y|x')^{s/\rho}\right)^{\rho}.
\end{align}
Through quantization of the channel inputs and outputs, one can evaluate these bounds for continuous channels as well.
Note that when $T=0$, we have that  $E_1(R,T,P_{x})=E_2(R,T,P_{x})$  and these two terms coincide also with Gallager error exponent, derived under the assumption of ML decoding~\cite[p.~210]{For68}.

As we shall see in Section \ref{sec:results}, the bounds
\eqref{eq:RCB_Pe}--\eqref{eq:RCB_Pu} are not accurate in the short-blocklength
regime, for the error probabilities typically considered in \ac{URLLC}.
Unfortunately, analyzing Forney's decoder using the probabilistic tools developed for the
short-blocklength regime is challenging \cite{Hayashi2015,Vincent2014}. To tackle this
issue, it will turn out convenient to use suboptimal decoders. We next introduce two
finite-blocklength achievability bounds that rely on such suboptimal decoders, as well as
on the \ac{RCU} bound~\cite[Thm. 16]{Polyanskiy2010}.
\subsection{Achievability Bound via CRC Outer Code}
\label{sec:DeltaBit}
We first consider an erasure decoder that relies on a CRC outer code for error detection.
The finite-blocklength performance of such a scheme was analyzed in~\cite{Jindal2011},
where the authors approximate the TEP $\pt$ using the so-called normal approximation~\cite[Eq.~(1)]{Polyanskiy2010}, and the UEP by $\pu \approx \pt 2^{-\Delta}$, where $\Delta$ is the number of parity bits added by the CRC outer code.
Next, we present a rigorous finite-blocklength achievability bound for this scheme.
\begin{thm}
	\label{thm:Delta}
	For an arbitrary input distribution $P_\rvecx$ and for every $\Delta \in \mathbb{N}_0$,
	there exists an $(n,k,\pt,\pu)$-code for the channel $P_{\rvecy|\rvecx}(\rvecy|\rvecx)$
	simultaneously satisfying $\pt\leq \pt\supp{ub,1}$ and  $\pu\leq \pu\supp{ub,1}$, where
	\begin{align}
		\label{eq:ptDelta_def}
		\pt\supp{ub,1} & = \mathrm{RCU}(k+\Delta,n)             \\
		\label{eq:puDelta_def}
		\pu\supp{ub,1} & = \mathrm{RCU}(k+\Delta,n) 2^{-\Delta}
	\end{align}
	with
	\begin{align}
		\label{eq:RCU_def}
		\mathrm{RCU}(k,n) & = \Exop_{}\mathopen{}\big[ \min\big\{1, (2^{k}-1) \nonumber                                                                          \\
		                  & \quad \times \Prob\ltrsqr{P_{\rvecy|\rvecx}(\rvecy|\bar{\rvecx}) \geq P_{\rvecy|\rvecx}(\rvecy|\rvecx) | \rvecx,\rvecy} \big\} \big]
	\end{align}
	and
	\begin{equation}
		\label{eq:xXbarYjoint}
		P_{\rvecx,\rvecy,\bar{\rvecx}}(\rvecx,\rvecy,\bar{\rvecx}) = P_{\rvecx}(\rvecx) P_{\rvecy|\rvecx}(\rvecy|\rvecx) P_{\rvecx}(\bar{\rvecx}).
	\end{equation}
\end{thm}
\begin{IEEEproof}
	See Appendix~\ref{appendix:proof_Delta}.
\end{IEEEproof}
\begin{rem}
	Note that, as we increase $\Delta$, the upper bound $\pu\supp{ub,1}$ on the
	undetected error probability $\pu$ decreases. At the same time, though, the coding
	rate $\rate=(k+\Delta)/n$ of the inner code increases, which causes $\pt\supp{ub,1}$ to
	increase as well. Also, for $\Delta = 0$, we have $\pt\supp{ub,1}=\pu\supp{ub,1}$, and
	the right-hand side of both \eqref{eq:ptDelta_def} and \eqref{eq:puDelta_def} reduce to
	$\mathrm{RCU}(k,n)$. Thus, we recover the RCU bound \cite[Thm. 16]{Polyanskiy2010}.
\end{rem}
\subsection{Achievability Bound via Generalized Information Density Thresholding}
\label{sec:Threshold}
We next state an achievability bound, which, inspired by \cite{Hayashi2015,Vincent2014}, is obtained by thresholding the so-called generalized information density~\cite{Martinez2011}.
Specifically, the decoder seeks the codeword with the highest likelihood and puts out the corresponding message if the generalized information density of this codeword (defined in \eqref{eq:GenInfoDens}) is higher than a preset threshold $n\lambda$. Otherwise, the decoder declares an erasure.

\begin{thm}
	\label{thm:Threshold_is}
	For an arbitrary input distribution $P_\rvecx$, for all $s~>~0$, and for all
	$\lambda \in \mathbb{R}$, there exists an $(n,k,\pt,\pu)$-code for the
	channel $P_{\rvecy|\rvecx}(\rvecy|\rvecx)$ satisfying $\pt\leq \pt\supp{ub,2}$ and
	$\pu\leq\pu\supp{ub,2}$, where
	\begin{align}
		\pt\supp{ub,2} & = \widetilde{\mathrm{RCU}}_{\lambda}(k,n) + \Prob\ltrsqr{\infden(\rvecx,\rvecy) < n \lambda}\label{eq:thm3-1} \\
		\pu\supp{ub,2} & = \Ex{}{\min\ltrcurley{1, (2^{k}-1) \widetilde{\psi}_{\vecy,\vecx}(\vecy,\vecx) }}.\label{eq:thm3-1b}
	\end{align}
	Here,
	\begin{equation}
		\label{eq:GenInfoDens}
		\infden(\vecx,\vecy) = \log_2 \frac{P_{\vecy|\vecx}(\vecy|\vecx)^s}{\Ex{\bar{\rvecx}}{P_{\vecy|\vecx}(\vecy|\bar{\rvecx})^s}}
	\end{equation}
	is the so-called generalized information density,
	\begin{align}
		 & \widetilde{\mathrm{RCU}}_{\lambda}(k,n)  = \Exop\mathopen{}\big[ \min\{1, (2^{k}-1) \nonumber                                                                                              \\
		 & \quad \times \Prob[P_{\rvecy|\rvecx}(\rvecy|\bar{\rvecx}) \geq P_{\rvecy|\rvecx}(\rvecy|\rvecx) | \rvecx,\rvecy ] \} \ind{\infden(\rvecx,\rvecy) \geq n \lambda}  \big]\label{eq:RCUtilde}
	\end{align}
	\begin{equation}
		\label{eq:Psi2Def}
		\widetilde{\psi}_{\vecy,\vecx}(\vecy,\vecx)\! = \!\Prob\ltrsqr{P_{\rvecy|\rvecx}(\rvecy|\bar{\rvecx})\! \geq \! \max\ltrcurley{\! P_{\rvecy|\rvecx}(\rvecy|\rvecx), \widetilde{\lambda}_{\vecy} \! }\!\! \ggiven \!\! \rvecx, \rvecy}
	\end{equation}
	\begin{equation}\label{eq:deltatilde}
		\widetilde{\lambda}_{\vecy} = \ltrp{2^{n\lambda} \Ex{\bar{\rvecx}}{P_{\rvecy|\rvecx}(\vecy|\bar{\rvecx})^{s}} }^{1/s}
	\end{equation}
	and $\rvecx$, $\bar{\rvecx}$ and $\rvecy$ are jointly distributed as in \eqref{eq:xXbarYjoint}.
\end{thm}
\begin{IEEEproof}
	See Appendix~\ref{appendix:proof_threshold_is}.
\end{IEEEproof}
\begin{rem}
	In Theorem~\ref{thm:Threshold_is}, $s>0$ is a parameter of the bound that can be optimized, similar to the input distribution $P_{\rvecx}$.
\end{rem}
\begin{rem}
	The optimal Forney's test \eqref{eq:forneytest}
	can equivalently be expressed as~\cite[Eq.~(15)]{For68}
	\begin{equation}
		\label{eq:Forney_rule2}
		P_{\vecx|\vecy}(\vecx|\vecy) > \frac{2^{nT}}{1+2^{nT}}.
	\end{equation}
	Using Bayes theorem and taking the logarithm of both sides, we may reformulate \eqref{eq:Forney_rule2} as
	\begin{equation}
		\label{eq:Forney_rule3}
		\log_2\frac{P_{\vecy|\vecx}(\vecy|\vecx)}{P_{\vecy}(\vecy)} > \log_2 \ltrp{2^k \frac{2^{nT}}{1+2^{nT}}}.
	\end{equation}
	Note that if we set $s=1$ in \eqref{eq:GenInfoDens}, we can see that the decoding rule
	employed in Theorem \ref{thm:Threshold_is} resembles Forney's rule. The key difference
	is that $P_{\rvecy}(\vecy)$ in \eqref{eq:Forney_rule3} is the output distribution
	induced by the chosen code, whereas
	$\Ex{\bar{\rvecx}}{P_{\rvecy|\rvecx}(\rvecy|\bar{\rvecx})}$, which is the denominator of
	the fraction in~\eqref{eq:GenInfoDens}
	for the case $s=1$, is the output distribution induced by the input distribution
	$P_{\rvecx}$.
\end{rem}
\begin{rem}
	If we let $\lambda \to -\infty$ in~\eqref{eq:thm3-1} and~\eqref{eq:deltatilde}, we
	recover the RCU bound~\cite[Thm.~16]{Polyanskiy2010} since
	$\lim_{\lambda\to-\infty}\widetilde{\mathrm{RCU}}_{\lambda}(k,n) =
		\mathrm{RCU}(k,n)$, and both $\pt\supp{ub,2}$ and $\pu\supp{ub,2}$ in~\eqref{eq:thm3-1}
	and~\eqref{eq:thm3-1b}, respectively, reduce to $\mathrm{RCU}(k,n)$.
\end{rem}

\subsection{Saddlepoint Approximation of Pairwise Error Probability}\label{sec:saddlepoint}
To numerically evaluate the bounds presented in Theorem~\ref{thm:Delta} and
Theorem~\ref{thm:Threshold_is}, one needs to evaluate a pairwise error probability
(namely~\eqref{eq:psiDef} in Appendix~\ref{appendix:proof_Delta} and \eqref{eq:Psi2Def})
with very high accuracy. Evaluating it via Monte-Carlo
averaging is time-consuming for low error probabilities. We next introduce a
saddlepoint approximation to evaluate the tail of the sum of independent but not
necessarily identically distributed random variables. Then, inspired by
\cite{font-segura18-03a}, we show that this saddlepoint approximation can be utilized to
evaluate \eqref{eq:psiDef} and \eqref{eq:Psi2Def} for
memoryless channels.

\begin{thm}
	\label{thm:SaddlePoint}
	Fix an $\omega \in \mathbb{R}$ and let $z_{i}$, $i \in \{1,\ldots,n\}$, be independent
	but not necessarily identically distributed random
	variables. Also, let $\gamma_i(\zeta)$ be the cumulant generating
	function (CGF) of $z_{i}$, and let $\gamma_i'(\zeta)$ and $\gamma_i''(\zeta)$ denote the first and
	second derivatives of $\gamma_i(\zeta)$, respectively.
	Let also $\gamma^{(n)}(\zeta) = \sum_{i=1}^{n} \gamma_i(\zeta)$, $\ltrp{\gamma^{(n)}(\zeta)}' = \sum_{i=1}^{n} \gamma'_i(\zeta)$ and $\ltrp{\gamma^{(n)}(\zeta)}'' = \sum_{i=1}^{n} \gamma''_i(\zeta)$.
	Suppose that there exists a
	$\zeta_{0}>0$ such that
	\begin{equation}\label{eq:fourth-moment-constraint}
		\sup_{\abs{\zeta}<\zeta_{0}} \abs{\odv[order=4]{\gamma_{i}(\zeta)}{\zeta}}<\infty, \quad \forall
		i\in\{1,\dots,n\}
	\end{equation}
	and also positive constants $m\sub{l}\leq m\sub{u}$ such that
	\begin{equation}\label{eq:second moment constraint}
		m\sub{l} \leq \sum_{i=1}^{n}\gamma_i''(\zeta) \leq m\sub{u}
	\end{equation}
	holds for all $n \in \mathbb{N}$ and for all $\abs{\zeta}\leq\zeta_{0}$.
	Assume that there exists a $\zeta \in [-\zeta_{0},\zeta_{0}]$ satisfying $\omega = \ltrp{\gamma^{(n)}(\zeta)}'$. If $\zeta > 0$, then
	\begin{align}
		\Prob\ltrsqr{ \sum_{i=1}^{n} z_{i} > \omega} & = e^{\gamma^{(n)}(\zeta) - \zeta \ltrp{\gamma^{(n)}(\zeta)}' + \frac{\zeta^2}{2} \ltrp{\gamma^{(n)}(\zeta)}''}  \nonumber \\
		                                             & \quad \times Q\ltrp{\zeta \sqrt{\ltrp{\gamma^{(n)}(\zeta)}''}} + o\ltrp{\frac{1}{\sqrt{n}}}.
		\label{eq:SaddlePoint1}
	\end{align}
	If $\zeta < 0$, then
	\begin{align}
		\Prob\ltrsqr{ \sum_{i=1}^{n} z_{i} > \omega} & = 1 - e^{\gamma^{(n)}(\zeta) - \zeta \ltrp{\gamma^{(n)}(\zeta)}' + \frac{\zeta^2}{2} \ltrp{\gamma^{(n)}(\zeta)}''}  \nonumber \\
		                                             & \quad \times Q\ltrp{-\zeta \sqrt{\ltrp{\gamma^{(n)}(\zeta)}''}}  + o\ltrp{\frac{1}{\sqrt{n}}}.
		\label{eq:SaddlePoint2}
	\end{align}
\end{thm}
\begin{IEEEproof}
	To prove Theorem~\ref{thm:SaddlePoint}, we follow the steps
	in~\cite[App. I]{Lancho2020}, with one crucial difference.
	Since the random variables $\rndz_{i}$ for $i \in
		\{1,\ldots,n\}$ are not necessarily identically distributed, we
	replace~\cite[Lem.~8]{Lancho2020} with~\cite[Thm.~1, Sec.~XVI.6]{Feller1971} in one of the steps detailed in~\cite[App. I]{Lancho2020}.
\end{IEEEproof}

For memoryless channels, the channel law can be factorized as
$P_{\rvecy|\rvecx}(\vecy|\rvecx) = \prod_{i=1}^{n} P_{\rndy|\rndx}(y_i|\rndx_i)$; thus,
$\log(P_{\rvecy|\rvecx}(\vecy|\rvecx)) = \sum_{i=1}^{n} \log(P_{y|\rndx}(y_i|\rndx_i))$.
We then obtain the desired saddlepoint approximation for \eqref{eq:psiDef}
by setting $z_{i} = \log(P_{y|\rndx}(y_i|\bar{\rndx}_i))$ and $\omega =
	\log(P_{\rvecy|\rvecx}(\vecy|\rvecx))$ for given $\rvecx$ and $\vecy$, and by omitting the $o(\cdot)$ terms in
\eqref{eq:SaddlePoint1} and \eqref{eq:SaddlePoint2}.
To obtain a saddlepoint approximation for~\eqref{eq:Psi2Def}, we set
instead
$\omega =
	\log\ltrp{\max\left\{P_{\rvecy|\rvecx}(\rvecy|\rvecx), \widetilde{\lambda}_{\vecy}
		\right\}}$.

The saddlepoint approximation in Theorem~\ref{thm:SaddlePoint} requires the evaluation
of the CGF $\gamma^{(n)}(\zeta)$ and its derivatives (or, equivalently, the
evaluation of $\gamma_i(\zeta)$ and its derivatives for all $i$). One may use Monte
Carlo averaging to evaluate $\gamma^{(n)}(\zeta)$ and then numerically evaluate
its derivatives; however, this is computationally expensive. Assuming we have an
i.i.d.~discrete codebook ensemble in which each symbol of every codeword is drawn
independently and uniformly from the finite-cardinality input set
$\setX$, we can evaluate $\gamma_i(\zeta)$ and its first and
second derivatives in closed form as
\begin{align}
	\gamma_i(\zeta)   & = \log \ltrp{\frac{1}{\Mi} \sum_{x \in \setX} e^{\zeta g_{\rndy_i}(x)} }                                                      \\
	\gamma_i'(\zeta)  & = \frac{\sum_{x \in \setX} g_{\rndy_i}(x) e^{\zeta g_{\rndy_i}(x)} }{\sum_{x \in \setX}  e^{\zeta g_{\rndy_i}(x)}}            \\
	\gamma_i''(\zeta) & = \frac{\sum_{x \in \setX} g_{\rndy_i}(x)^2 e^{\zeta g_{\rndy_i}(x)} }{\sum_{x \in \setX} e^{\zeta g_{\rndy_i}(x)}} \nonumber \\  &\quad- \ltrp{\frac{\sum_{x \in \setX} g_{\rndy_i}(x) e^{\zeta g_{\rndy_i}(x)} }{\sum_{x \in \setX} e^{\zeta g_{\rndy_i}(x)}} }^2
\end{align}
where
\begin{equation}
	g_{y_i}(x) = \log\ltrp{P_{\rndy|\rndx} (y_i|x)}.
\end{equation}

%%%%%%%%%%%%%%%%%%%%%%%%%%%%%%%%%%%%%%%%%%%%%%%%%%%%%%%%%%%%%%%%%%%%%%%%%%%%%%%%%%%%%%%%%%%%%%%%%%%%%%%%%%%%%%%%%%%%%%%%%%%%%%%%%%
\section{Error Detection for CA Polar Codes}\label{sec:error_det}
We next discuss the design of error detection and decoding strategies for CA
	polar codes.
	We start by considering \ac{SCL} decoding of polar codes~\cite{Tal2013} as a reference
	scheme, and we
	highlight its basic error detection capability, which enables one to trade in a very limited way a reduction of the UEP with an increase of the TEP.
	Inspired by Theorems~\ref{thm:Delta} and~\ref{thm:Threshold_is}, we then introduce two
	decoding strategies that provide much more flexibility.
	Specifically, similar to Theorem~\ref{thm:Delta}, the first strategy relies on the use
	of an outer code for error detection.
	In CA \ac{SCL} polar codes, this can be achieved by reserving a subset of the bits of
	the already existing \ac{CRC} code for error detection.
	Similar to Theorem~\ref{thm:Threshold_is}, the second strategy relies on a threshold
	test.
	Specifically, we consider Forney's optimal test~\eqref{eq:forneytest}, applied to the
	list of codewords returned by the \ac{SCL} decoder. This allows us to approximate the
	metric in~\eqref{eq:forneytt}.

\begin{figure*}[t]
	\centering
    \includegraphics[width = 0.8\textwidth]{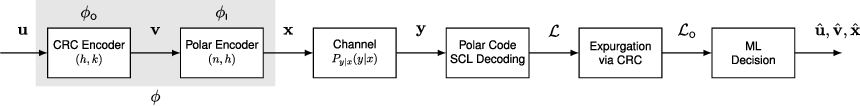}
	\caption{Reference model describing the encoding with a CA polar code, transmission over the channel, and \ac{SCL} decoding.}
	\label{fig:setup}
\end{figure*}

\subsection{CA Polar Codes: SCL Decoding and Error Detection}
\label{sec:polar:basic}

The reference scheme is based on the serial concatenation of an inner polar code $\incode$
with an outer \ac{CRC} code $\outcode$ \cite{Tal2013}, and it is depicted in
Fig.~\ref{fig:setup}. At the encoder side, the binary representation $\bm{u} \in \fieldtwo^k$
	of an arbitrary message $w\in\{1,\dots, 2^{k}\}$ is encoded
via an $(\outN,k)$ systematic \ac{CRC} encoder, which appends $\Delta = \outN-k$ parity
bits to the input message. The output of the \ac{CRC} encoder is denoted by $\bm{v} \in
		\fieldtwo^{h}$, and
it is provided as input to a polar code encoder. We denote by $\bm{x}$ the polar code
encoder output, which we assume to be matched to the channel input alphabet
$\setX$. That is, denoting by $n_c$ the blocklength in bits of the polar code, we have that $\bm{x}
	\in \setX^n$ where $n = n_c / \log_2 \Mi$. The outer-code rate is $\outR = k/h$, the
inner-code rate is $\inR = h/n_c$, and the overall code rate, expressed in information
bits per channel use, is $R = k/n = \inR \outR\log_2 \Mi$.
As in Section~\ref{sec:SystemModel}, we denote by  $\code \subset \setX^n$ the overall code
	defined by this concatenation.
We
denote the outer \ac{CRC} encoder and the inner polar code encoder as
\begin{equation}
	\outENC: \fieldtwo^k \mapsto \fieldtwo^h \quad \text{and} \quad \inENC: \fieldtwo^h \mapsto \setX^n
\end{equation}
respectively,
whereas, again in agreement with Section~\ref{sec:SystemModel}, the overall encoding function is $\ENC = \inENC\, \circ \, \outENC$.
Following \cite{Tal2013}, we assume that \ac{SCL} decoding of the polar code is performed.
This yields a list of codewords $\plist \subseteq \fieldtwo^h$.
Throughout, we shall denote by $L$ the list size.
The list is then expurgated by removing all its elements that do not satisfy the outer code constraints, resulting in the list $\olist$. Using the shorthand
\begin{equation}
	\inENC(\olist) = \{\bm{x}' \in \setX^n | \bm{x}' = \inENC(\bm{v}'), \bm{v}' \in \olist  \}
\end{equation}
we can write the final decision, which follows by applying the \ac{ML}
decision criterion to the list $\olist$, as
\begin{equation}
	\hat{\bm{x}} = \argmax_{\bm{x'} \in \inENC(\olist)} P_{\bm{y}|\bm{x}}(\bm{y}|\bm{x}'). \label{eq:ML_in_list}
\end{equation} 
Specifically, the overall decoder returns the message corresponding to the
codeword $\hat{\vecx}$ if $\olist$ is non-empty. Otherwise, it declares an
erasure.

\begin{rem}\label{rem:SCL}
	This decoding algorithm provides only a very limited error detection capability:
	an erasure is declared whenever $\olist$ is empty.
	For a given
	\ac{CRC} code and a given polar code, the only degree of freedom at our disposal to
		control the trade-off
	between \ac{TEP} and \ac{UEP} is the choice of the \ac{SCL} decoder list size $L$.
	A large $L$ results in a \ac{TEP} that is close to the one attainable by applying
		the \ac{ML} decoding rule to the overall code.
	However, the larger $L$, the smaller the probability that $\olist$ will be empty; hence,
	the higher the \ac{UEP}.\footnote{Observe that in the extreme case $L=2^{h}$, the list
	$\olist$ is always nonempty and \ac{TEP} and \ac{UEP} coincide.}
	Conversely, the smaller $L$, the higher the probability that $\olist$ will be
		empty, and, hence, the smaller the \ac{UEP}, at the expenses of an increased
		\ac{TEP}.
\end{rem}

\subsection{SCL Decoding with Split CRC \emph{(Algorithm \texttt{A})}}\label{subsec:split}
\begin{figure}[t]
	\centering
    \includegraphics[width = 0.83\columnwidth]{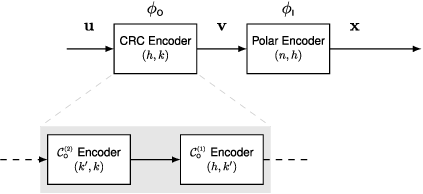}
	\caption{Decomposition of the outer CRC code $\outcode$ as the concatenation of two binary linear block codes.}
	\label{fig:setup_algB}
\end{figure}

As already mentioned, our first strategy to increase our ability to trade \ac{TEP}
	with \ac{UEP} relies on the use of an outer \ac{CRC} code for error detection, similar
	to Theorem~\ref{thm:Delta}.
Specifically, we decompose the \ac{CRC} code that is already present in CA polar
	codes in the concatenation of two binary linear block codes $\outcodex{1}$ and
	$\outcodex{2}$. The code $\outcodex{1}$ has parameters $(h,k')$, and code $\outcodex{2}$
	has parameters $(k',k)$. Under systematic encoding, the second code appends $\Delta_2 =
		k'-k$ parity bits to the binary representation $\mathbf{u}$ of the information message.
	The output of the second code encoder is used as input for the encoder of the first code,
	which appends additional $\Delta_1 = h - k'$ bits to the second encoder output. The
	overall number of parity bits is $\Delta = \Delta_1 + \Delta_2$. By suitably choosing
	$\outcodex{1}$ and $\outcodex{2}$, we can ensure that the concatenation of the two codes is equivalent to the
	outer code $\outcode$, resulting in an unmodified encoding function $\outENC$. The
	concatenation is illustrated in Figure~\ref{fig:setup_algB}.

At the decoder side, the first outer code $\outcodex{1}$ is used to expurgate the
	list $\plist$ produced by the polar \ac{SCL} decoder. That is, the concatenation of
	$\outcodex{1}$ with the inner polar code $\incode$ is treated as an $(n,k')$ CA polar
	code. We denote by $\onelist$ the list at the output of the \ac{SCL} decoder,
	after the expurgation performed via  $\outcodex{1}$, i.e. $\onelist = \plist \cap \outcodex{1}$. Decoding then proceeds as follows.

\medskip

\begin{itemize}
	\addtolength{\itemindent}{2em}
	\setlength{\itemsep}{1.5ex}
	\item[\emph{Case 1:}] $\onelist$ is empty. The decoder returns an erasure.
	\item[\emph{Case 2:}] $\onelist$ is non-empty. The decoder computes
	      \begin{equation}
		      \hat{\bm{x}} = \argmax_{\bm{x'} \in \inENC(\onelist)} P_{\bm{y}|\bm{x}}(\bm{y}|\bm{x}') \label{eq:ML_in_list2}.
	      \end{equation}
	      If $\hat{\bm{x}}$ satisfies the $\Delta_2$ parity-check constraints imposed
	      by the second outer code $\outcodex{2}$, then the decoder outputs the message
		      corresponding to $\hat{\bm{x}}$.
	      Otherwise, the decoder rejects the decision and outputs an erasure.
\end{itemize}

\medskip

In the following, we will refer to the decision rule described above as Algorithm
\texttt{A}. It is important to stress that, in Algorithm \texttt{A}, part of
	the redundancy introduced by the outer code $\outcode$ is used exclusively for error detection.
	Specifically, the second outer code $\outcodex{2}$ is used as an error detection code,
	applied to the $(n,k')$ CA polar code formed by the concatenation of $\outcodex{1}$ with
	$\incode$.

\begin{rem}
	Note that Algorithm \texttt{A} reduces to the reference scheme described in
	Section \ref{sec:polar:basic} whenever $\Delta_{2}=0$ and, hence, $\Delta_1 = \Delta$. In
	contrast, for $\Delta_1 = 0$, we have that $\onelist = \plist$, which implies that
	all \ac{CRC} code constraints are dedicated to error detection.
	For a fixed list size $L$, this results in a much lower \ac{UEP} than the reference
		scheme, at the cost of a higher \ac{TEP}.
	Intermediate values of
	$\Delta_1$ can be
	used to achieve a different tradeoff between the \ac{TEP} and the \ac{UEP}.
\end{rem}

\subsection{\ac{SCL} Decoding with Threshold Test \emph{(Algorithm \texttt{B})}}\label{subsec:SCLTT}
Similar to Theorem~\ref{thm:Threshold_is}, our second approach relies on a threshold
	test. Specifically, we modify the SCL algorithm of the reference scheme described in
	Section~\ref{sec:polar:basic} by introducing an additional error detection mechanism in the form of a threshold test. Namely, upon obtaining the
	expurgated list $\olist$, the decoder operates as follows:

\medskip

\begin{itemize}
	\addtolength{\itemindent}{2em}
	\setlength{\itemsep}{1.5ex}
	\item[\emph{Case 1:}] $\olist$ is empty. The decoder returns an erasure.
	\item[\emph{Case 2:}] $\olist$ contains a single element $\bm{v}'$. The decoder outputs
	      the message corresponding to
	      \begin{equation}
		      \hat{\bm{x}} = \inENC(\bm{v}').
	      \end{equation}
	\item[\emph{Case 3:}] $\olist$ contains more than one element. The decoder computes a preliminary decision according to \eqref{eq:ML_in_list}. A threshold test is then performed by computing Forney's test~\eqref{eq:forneytt} on the codewords in $\inENC(\olist)$:
	      \begin{equation}
		      \SCLTT(\bm y, \hat{\bm x}) = \frac{P_{\bm{y}|\bm{x}}\left(\bm y |\hat{\bm x}\right)}{\sum_{\bm{x'} \in \inENC(\olist)\setminus\{\hat{\bm x}\}}P_{\bm{y}|\bm{x}}\left(\bm y |\bm{x}'\right)}. \label{eq:SCLtt}
	      \end{equation}
	      The decision $\hat{\bm{x}}$ is accepted if
	      \begin{equation}
		      \SCLTT(\bm y, \hat{\bm x}) \geq 2^{nT} \label{eq:SCLtest}
	      \end{equation}
	      and it is rejected otherwise, resulting in an erasure.
\end{itemize}

\medskip

In the following, we will refer to the decision rule described above as Algorithm \texttt{B}.

\begin{rem}
	Note that in case $1$ and case $2$ above, the algorithm operates exactly as in
		the reference scheme reviewed in Section~\ref{sec:polar:basic}.
	The difference resides in case $3$, where the reference scheme returns the message
		corresponding to the codeword in $\inENC(\olist)$ that maximizes the likelihood.
		Instead, Algorithm \texttt{B} relies on a threshold test that approximates
	Forney's metric \eqref{eq:forneytt} by replacing the sum over the codebook with the sum
	over the expurgated list $\inENC(\olist)$. 
	In the limiting case $L=2^{h}$, the metric in~\eqref{eq:SCLtt} reduces to
	Forney's metric~\eqref{eq:forneytt}, which however is not computable for CA polar codes
	for values of $h$ of practical interest. To summarize, the proposed strategy provides an
	effective method to implement Forney's rule in CA polar codes.
\end{rem}

%%%%%%%%%%%%%%%%%%%%%%%%%%%%%%%%%%%%%%%%%%%%%%%%%%%%%%%%%%%%%%%%%%%%%%%%%%%%%%%%%%%%%%%%%%%%%%%%%%%%%%%%%%%%%%%%%%%%%%%%%%%%%%%%%%
\section{Numerical Results}\label{sec:results}

In this section, we report numerical results to investigate the accuracy of the
finite-blocklength bounds proposed in Theorems~\ref{thm:Delta}
and~\ref{thm:Threshold_is} in comparison to Forney's
bound~\eqref{eq:RCB_Pe}--\eqref{eq:RCB_Pu}, as well as the performance of the error
detection methods proposed in Section~\ref{sec:error_det} for CA polar codes, in
relation to the proposed bounds.
We also aim to understand when threshold-based schemes should be preferred over
solutions based on outer CRC codes.
To achieve these goals, we consider two channel models:
a \ac{biAWGN} channel, and a block-memoryless phase-noise channel~\cite{Peleg98} with pilot-aided phase estimation, \ac{QPSK} modulation, and mismatched decoding.
We also assume that all codewords $\vecx_m$, $m=1,\dots, 2^k$, in Definition~\ref{def:code} are subject to the power constraint $\vecnorm{\vecx_m}^2=n$.

The results of CA polar codes are obtained via Monte Carlo simulations. In all
simulation results, the polar codes have been designed by selecting the indexes of the $h$ information
bits that feature the largest mutual information under genie-aided \ac{SC} decoding
\cite{Ari09}.
This mutual information is determined via a \ac{DE} analysis \cite{Mori2009,Tal2013,RU01a} that relies on a Gaussian
approximation \cite{chung2001analysis,Tri12}. To compare the performance of the various
decoding algorithms with the achievability bounds presented in Section~\ref{sec:bounds},
we fix a \emph{target total error probability} $\pt^\star$ and a \emph{target undetected
	error probability} $\pu^\star$, and we compute the lowest \ac{SNR} for which
the following two inequalities hold: $\pt \leq
	\pt^\star$ and $\pu \leq \pu^\star$. Specifically, denoting by $E_b$ the energy per
information bit, and by $N_0$ the single-sided \ac{AWGN} power spectral density, we make
use of the following definition:

\begin{dfn}[SNR threshold]\label{def:SNRthreshold}
	Given a blocklength $n$, a rate $R$, and a channel \ac{SNR} $E_b/N_0$, denote by
	$\mathcal{P}_{n,R}(E_b/N_0)$ the set of  achievable pairs $(\pt,\pu)$. For the target error probabilities $\pt^\star$ and $\pu^\star$, the \ac{SNR} threshold is defined as
	\begin{equation}
		\gamma(\pt^\star,\pu^\star) = \min \left\{\frac{E_b}{N_0} \bigg| \left(\pt^\star, \pu^\star \right) \in \mathcal{P}_{n,R}(E_b/N_0) \right\}. \label{eq:gamma}
	\end{equation}
\end{dfn}
For each coding scheme, the reported $\gamma(\pt^\star,\pu^\star)$ is optimized
over the parameters of the decoding algorithm. Specifically, for Algorithm~\texttt{A},
the optimization involves finding the best split between $\Delta_1$ and $\Delta_2$,
whereas, for Algorithm~\texttt{B}, it involves the threshold $T$ in~\eqref{eq:SCLtest}.

We also evaluate an upper bound on $\gamma(\pt^\star,\pu^\star)$ by means of the two achievability bounds in
Section~\ref{sec:bounds}, computed by considering an input distribution for which each symbol in
each codeword is drawn independently and uniformly from the constellation set.
Specifically, given two target \ac{TEP} and
\ac{UEP} values $\pt^{\star}$ and $\pu^{\star}$ and a given \ac{SNR}, we set $\Delta =
	\ceil{\log(\pt^{\star}/\pu^{\star})}$  in Theorem~\ref{thm:Delta}.
Then, we use~\eqref{eq:ptDelta_def} to evaluate $\pt\supp{ub,1}$.
If $\pt\supp{ub,1}>\pu^{\star}$ we increase the \ac{SNR}; otherwise we lower it. This process
is repeated until convergence.
In the case of Theorem~\ref{thm:Threshold_is}, we optimize $\lambda$ and $s$ so that
$\pu\supp{ub,2}$ in~\eqref{eq:thm3-1b} coincides
with $\pu^\star$. Then we evaluate the corresponding $\pt\supp{ub,2}$ in~  \eqref{eq:thm3-1}. If
$\pt\supp{ub,2}>\pt^{\star}$, we increase the SNR. Otherwise, we lower it. This process is
repeated until convergence.
The pairwise error probabilities that appear in the bounds in
Theorem~\ref{thm:Delta} and Theorem~\ref{thm:Threshold_is} are computed via the
saddlepoint approximation provided in Theorem~\ref{thm:SaddlePoint}.\footnote{Specifically, for the biAWGN case, we use the numerically efficient implementation of the saddlepoint approximation proposed in~\cite{font-segura18-03a}, whereas the procedure described in Section~\ref{sec:saddlepoint} is used for the block-memoryless phase-noise channel. The code used to plot the bounds can be found at \url{https://github.com/OguzKislal/ErrorDetection_InfoTheory}.}
In all the results that follow, we set $\pt^\star = 10^{-3}$
and $\pu^\star = 10^{-5}$.
%%--------------------------------------------------------------------------------------
\subsection{Results for the Binary-Input AWGN Channel}\label{sec:results:biAWGN}
We consider \ac{BPSK} transmission under the model
\begin{equation}
	y_i = x_i + z_i, \quad i=1,\ldots,n.
\end{equation}
Here, $x_i \in \{+1, -1\}$ and the $z_i$ are independent and identically distributed (\iid), and follow a Gaussian distribution with zero mean and variance $\sigma^2$. For this channel model, we have $E_b/N_0 = 1/(2R\sigma^2)$.

\begin{figure}[t]
	\centering
	\includegraphics[width = 0.95\columnwidth]{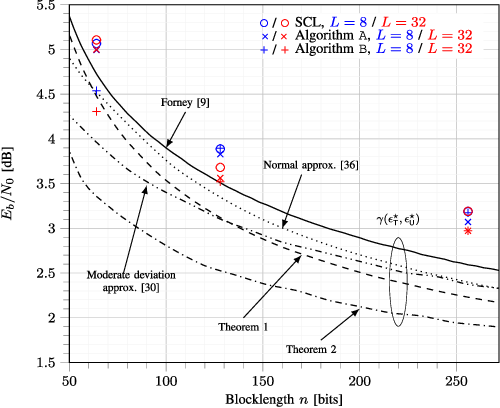}
	\caption{The minimum ${E_b}/{N_0}$ to achieve $\pt^{\star}=10^{-3}$ and $\pu^{\star}=10^{-5}$ as a function of blocklength $n$. Here, we consider a biAWGN channel with $R=0.5$ bits per channel use.}
	\label{fig:ForneyBound}
\end{figure}

In Fig.~\ref{fig:ForneyBound}, we consider the case $R=1/2$ bits per channel use, and
depict the minimum \ac{SNR} threshold $\gamma(\pt^*,\pu^*)$ as a function of the
blocklength.
Specifically, we illustrate three upper bounds on
$\gamma(\pt^*,\pu^*)$, obtained via Forney's bound in
\eqref{eq:RCB_Pe}--\eqref{eq:RCB_Pu} and via the two
achievability bounds introduced in Theorem~\ref{thm:Delta}
and~\ref{thm:Threshold_is}, respectively.\footnote{The expectations required to
	evaluate~\eqref{eq:RCU_def},~\eqref{eq:thm3-1},~\eqref{eq:thm3-1b},
	and~\eqref{eq:RCUtilde} are evaluated via Monte Carlo simulations.}
For completeness, we also report two approximations: the one based on the normal
approximation proposed in~\cite{Jindal2011} (which approximates the result
provided in Theorem~\ref{thm:Delta}) and the one based on a moderate-deviation analysis,
which follows from~\cite[Thm.~1]{Hayashi2015} and relies on a threshold decoded similar
to the one used in
Theorem~\ref{thm:Threshold_is}.
Three CA polar codes are also considered: a $(64,32)$ CA polar code based on a $6$-bit
\ac{CRC} code (i.e., $\Delta=6$)  with polynomial $\mathtt{0x43}$; a $(128,64)$ CA polar
code based on a $7$-bit \ac{CRC} code ($\Delta=7$)  with polynomial $\mathtt{0x89}$; and a
$(256,128)$ CA polar code based on a $8$-bit \ac{CRC} code ($\Delta=8$)  with polynomial
$\mathtt{0x1D5}$. \ac{SCL} decoding with two list sizes ($L=8$ and $L=32$) is considered
in the simulations.

We note that the achievability bounds introduced in Theorem \ref{thm:Delta} and
\ref{thm:Threshold_is} provide a much lower estimate of the minimum \ac{SNR}
threshold than Forney's bound.
Furthermore, the achievability
bound based on generalized information density thresholding (Theorem
\ref{thm:Threshold_is}) provides the lowest estimate.\footnote{As we will show in
	Fig.~\ref{fig:sVsSNR_ErrProb},  optimizing over the parameters $s$ is crucial to get such
	low estimate.}
Note that the two approximations reported in the figure are not
particularly accurate and tend to
overestimate the SNR threshold.
We also observe that, at very short blocklengths, the gap between the
achievability bounds in Theorem~\ref{thm:Delta} and Theorem~\ref{thm:Threshold_is} is the
largest: for $n=64$, the gap is \qty{1.1}{dB}; it reduces to \qty{0.3}{dB} when $n=264$.
This result suggests that using a threshold to detect errors is preferable to using
an outer \ac{CRC} code when $n$ is small.
The intuition is that, in detection schemes based on an outer \ac{CRC} code,
the inner code rate
needs to be increased to compensate for the addition of \ac{CRC} bits.
This is, however, detrimental when $n$ is small.

The results obtained with CA polar codes confirm this observation: the threshold approach used in Algorithm~\texttt{B} yields a significant gain over the \ac{CRC} based approach used in Algorithm~\texttt{A} when $n=64$.
Specifically, the gain is approximately \qty{0.5}{dB} for
$L=8$, and it increases to \qty{0.7}{dB} for $L=32$.
Interestingly, Algorithm $\texttt{A}$ yields only a
limited gain (between $0.1$ and $\qty{0.2}{dB}$) over the basic error detection
capability provided by plain \ac{SCL} decoding of CA polar codes (see Remark
\ref{rem:SCL}). Remarkably, Algorithm~\texttt{B} allows one to operate below
Forney's achievability bound, as well as below the CRC-based achievability bound given in
Theorem~\ref{thm:Delta}, and to operate less than \qty{1}{dB} away from the threshold-based achievability bound
of Theorem~\ref{thm:Threshold_is}.
In agreement with the behavior of the underlying bounds, as the blocklength grows, the gap between the different
strategies diminishes. When $n=128$, Algorithm~\texttt{B} still yields the best performance for all the considered list sizes, but it is closely matched by the performance of
Algorithm \texttt{A}. When $n=256$, Algorithm~\texttt{A} becomes very competitive: with a list size $L=32$, its performance is the same
as that of Algorithm \texttt{B}; with a list size $L=8$, it even outperforms Algorithm~\texttt{B}, although slightly.  This behavior may be explained as follows: as
the blocklength grows, the rate penalty introduced by the outer \ac{CRC} diminishes, and
\ac{CRC}s with a larger number of parity bits can be used. This allows for the exploration
of a larger set of pairs $(\Delta_1,\Delta_2)$, which enables a finer tuning of the error
detection capability provided by the algorithm.

\begin{figure}[t]
	\centering
	\includegraphics[width = 0.95\columnwidth]{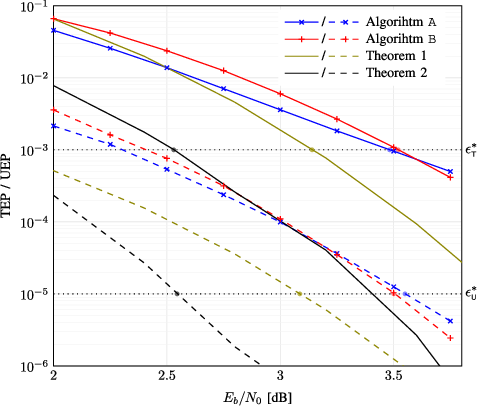}
	\caption{\ac{TEP} and \ac{UEP} versus ${E_b}/{N_0}$ for $(128, 64)$ for CA polar codes under Algorithm~\texttt{A} and Algorithm~\texttt{B} ($L=32$). The bounds of Theorem 1 and Theorem 2 are included as a reference. Solid lines are used for the \ac{TEP}, while dashed lines are used for the \ac{UEP}. The target \ac{TEP} and \ac{UEP} used for the optimization of the error detection parameters are $\pt^* = 10^{-3}$ and $\pt^* = 10^{-5}$.}
	\label{fig:FER_UER}
\end{figure}

In Fig.~\ref{fig:FER_UER}, we study the behavior of the \ac{TEP} and \ac{UEP} as a
function of ${E_b}/{N_0}$ for $n=128$ and $R=0.5$. All curves are computed for
parameters corresponding to the ones that minimize the \ac{SNR} threshold for the target
$\pt^* = 10^{-3}$ and $\pu^* = 10^{-5}$.
For the CA polar code,  we use \ac{SCL} decoding
with list size $L = 32$. Consistently with the results already presented in
Fig.~\ref{fig:ForneyBound}, Algorithm~\texttt{A} attains the target error probability at
${E_b}/{N_0} \approx 3.55~\dB$ and Algorithm~\texttt{B} attains it at ${E_b}/{N_0} \approx
	3.52~\dB$. We also report the bounds given in Theorem~\ref{thm:Delta} and
Theorem~\ref{thm:Threshold_is} as a reference. Similarly to the result in
Fig.~\ref{fig:ForneyBound}, the achievability bound obtained through
Theorem~\ref{thm:Threshold_is} shows that, for the biAWGN channel, performance can be
significantly improved using a decoding strategy based on thresholding the generalized
information density.

\begin{figure}[t]
	\centering
	\includegraphics[width = 0.95\columnwidth]{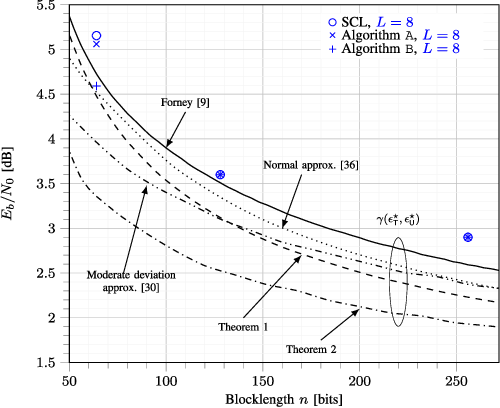}
	\caption{The minimum ${E_b}/{N_0}$ to achieve $\pt^{\star}=10^{-3}$ and $\pu^{\star}=10^{-5}$ as a function of blocklength $n$. The CRC polynomials and the  frozen bit set are chosen according to~\cite{3GPP21}. A CRC-$6$ is used for $n=64$, while a CRC-$11$ is used for $n=128,256$. Here, we consider a biAWGN channel with $R=0.5$ bits per channel use.}
	\label{fig:ForneyBound_5G}
\end{figure}

In Fig.~\ref{fig:ForneyBound_5G}, the performance achievable by using the CA polar
	codes of the $5$G NR standard~\cite{3GPP21} is reported for the case of $R=0.5$. The $5$G
	NR standard offers a variety of \ac{CRC} code polynomials, ranging from a $6$-bit \ac{CRC}
	code to a $24$-bit \ac{CRC} code. In the analysis, the $6$-bit \ac{CRC} code has been used for the blocklength $n=64$, and the $11$-bit \ac{CRC} has been used for the blocklenghts $n=128$ and $n=256$. This selection allows minimizing the \ac{SNR} thresholds achieved by the $5$G NR codes for the target error probabilities. In the
	simulations,  the list size has been restricted to $L=8$. The performance achieved by the $5$G NR codes is comparable to the one for the codes provided in  Fig.~\ref{fig:ForneyBound}. It is possible to observe that, for the two largest blocklengths, the use of a $11$-bit \ac{CRC} provides an error detection capability that exceeds the one required by our
	optimization target ($\pt^* = 10^{-3}$ and $\pu^* = 10^{-5}$), even under standard \ac{SCL} decoding. Hence, neither Algorithm \texttt{A} nor Algorithm \texttt{B} can provide a gain in this setting. The result may change by setting lower undetected error probability targets.

\begin{figure}[t]
	\centering
	\includegraphics[width = 0.95\columnwidth]{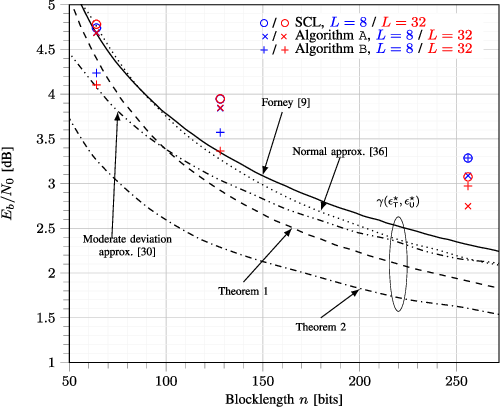}
	\caption{The minimum ${E_b}/{N_0}$ to achieve $\pt^{\star}=10^{-3}$ and $\pu^{\star}=10^{-5}$ as a function of blocklength $n$. Here, we consider a biAWGN channel with $R\simeq 0.33$ bits per channel use.}
	\label{fig:ForneyBoundLowRate}
\end{figure}

In Fig.~\ref{fig:ForneyBoundLowRate}, we consider the case $R\simeq 1/3$.
Again, CA polar codes designed for three blocklengths ($n=64$, $128$, and $256$) are considered for the
simulations. The \ac{CRC} codes used in the three cases are the same as
the ones used
for the rate-$1/2$ setting of Fig.~\ref{fig:ForneyBound}.
We see from the figure that most of the insights discussed for the rate-$1/2$ case extend
to the rate-$1/3$ case, with one notable difference.
When $R\simeq 1/3$, Algorithm~\texttt{B} clearly outperforms
Algorithm~\texttt{A} when $n=64$ and $n=128$, whereas Algorithm~\texttt{A} is clearly preferable when $n=256$.

\begin{figure}[t]
	\centering
	\includegraphics[width = 0.95\columnwidth]{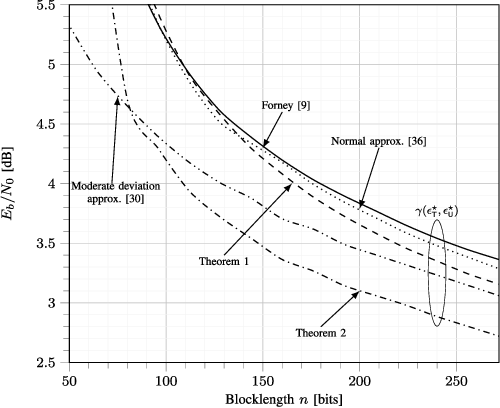}
	\caption{The minimum ${E_b}/{N_0}$ to achieve $\pt^{\star}=10^{-6}$ and $\pu^{\star}=10^{-9}$ as a function of blocklength $n$. Here, we consider a biAWGN channel with $R=0.5$ bits per channel use.}
	\label{fig:ForneyBound_lowEps}
\end{figure}

In Fig.~\ref{fig:ForneyBound_lowEps}, we consider the case $R=0.5$ but much lower target
\ac{TEP} and \ac{UEP}. Specifically, we set
$\pt^{\star}=10^{-6}$ and $\pu^{\star}=10^{-9}$. For this setting, we present only our
achievability bounds since the simulation of polar codes at these low error values is
challenging.
As shown in the figure, for these parameters, Forney's bound provides an estimate of the
minimum SNR that is similar to the one provided by Theorem~\ref{thm:Delta}. However, the estimate
provided by Theorem~\ref{thm:Threshold_is} is still much more accurate.

\subsection{Results for the Block-Memoryless Phase-Noise Channel}\label{sec:results:phasenoise}

It is worth highlighting that Forney's optimal threshold test \eqref{eq:forneytt},
the thresholding scheme used in Theorem~\ref{thm:Threshold_is}, and the test~\eqref{eq:SCLtt}
performed by Algorithm~\texttt{B} all require precise knowledge of the channel law
$P_{\bm{y}|\bm{x}}$. In contrast, the error detection mechanism used in both the
achievability bound in Theorem~\ref{thm:Delta} and by Algorithm~\texttt{A}
relies solely on a \ac{CRC} test. It is, therefore, interesting to study how the
insights gained via the analysis discussed in Section~\ref{sec:results:biAWGN} change in the presence
of an imprecise (noisy) knowledge of the channel law, which may originate, for instance, from
imperfect \ac{CSI}. To perform this analysis, we consider the block-memoryless phase-noise
channel \cite{Peleg98}---a model that is relevant, e.g., for satellite uplink channels
with time division multiple access \cite{Liva2013:Lowrate_TCOM}. Specifically, the system
relies on \ac{QPSK} transmission under the model
\begin{equation}
	y_i = e^{j\theta} x_i + z_i \quad i=1,\ldots,n.
\end{equation}
Here, $x_i \in \left\{\pm {\sqrt{2}}/{2} \pm j{\sqrt{2}}/{2} \right\}$ and the $z_i$  are
i.i.d.~and follow a complex Gaussian distribution with zero mean and variance $2\sigma^2$. For this channel model, $E_b/N_0 = 1/(2R\sigma^2)$. The phase rotation $\theta$ is
constant over the packet transmission, and it is drawn independently and uniformly at
random from the interval $[0,2\pi)$ at each packet transmission. We assume that we can
estimate the noise variance perfectly (as it does not change from packet to packet).
On the contrary, we assume that the receiver has only an estimate $\hat{\theta}$ of the phase rotation~$\theta$. Indeed, $\theta$ needs to be estimated for each received
packet and this may result in a non-negligible estimation error.
The receiver proceeds by computing the mismatched likelihoods
\begin{equation}
	q(y_i,x_i,\hat\theta) = \frac{1}{2 \pi \sigma^2} \exp \left(- \frac{1}{2\sigma^2} \left|y_i - e^{-j\hat\theta}x_i\right|^2 \right)
\end{equation}
for $i=1,\ldots,n$, which are then provided as input to the decoder.
For the analysis, we consider the case where $\hat\theta$ is obtained via \ac{ML}
estimation performed over a pilot field composed by $\npilot$ \ac{QPSK}
symbols.\footnote{In the results that follow, the energy overhead entailed by the transmission of pilot
	symbols is neglected: it would only cause a rigid shift of the $E_b/N_0$, which applies
	to all simulation results and achievability bounds.}
Note that the achievability bounds in Theorem~\ref{thm:Delta} and~\ref{thm:Threshold_is} can be evaluated for this setting by replacing
$P_{\rvecy|\rvecx}(\rvecy|\rvecx)$ with $\prod_{i=1}^{n} q(\rndy_i,\rndx_i,\hat{\theta})$
and by averaging the bounds over $\theta$ and $\hat{\theta}$.

\begin{figure}[t]
	\centering
	\includegraphics[width = 0.95\columnwidth]{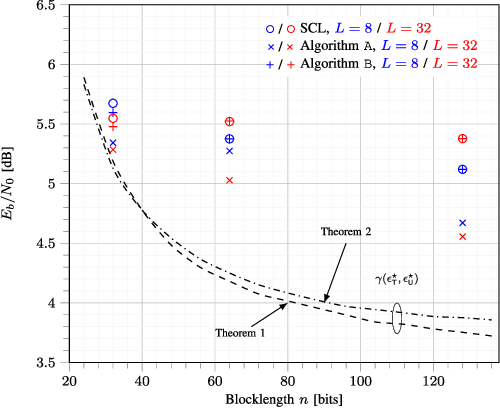}
	\caption{The minimum ${E_b}/{N_0}$ to achieve $\pt^{\star}=10^{-3}$ and $\pu^{\star}=10^{-5}$ as a function of blocklength $n$. Here, we consider QPSK modulated symbols transmitted over a block-memoryless phase-noise channel channel with 10 pilot symbols and $R=1$ bits per channel use.}
	\label{fig:Mismatch}
\end{figure}

In Fig.~\ref{fig:Mismatch}, we report the minimum \ac{SNR} required to achieve the target
error probability pair $\pt^\star = 10^{-3}$ and $\pu^\star = 10^{-5}$, as a function of
blocklength $n$. The CA polar codes used for the simulations are the same as the one
considered in
Fig.~\ref{fig:ForneyBound}. Since these codes have rate $1/2$ and since we use QPSK
modulation, the overall rate is $R = 1$ bits per channel use. To evaluate the robustness to \ac{CSI}
mismatch of the different error detection strategies, we consider a setting where the
phase estimate is obtained using only $\npilot=10$ pilot symbols.

\begin{figure}
	\centering
	\includegraphics[width = 0.95\columnwidth]{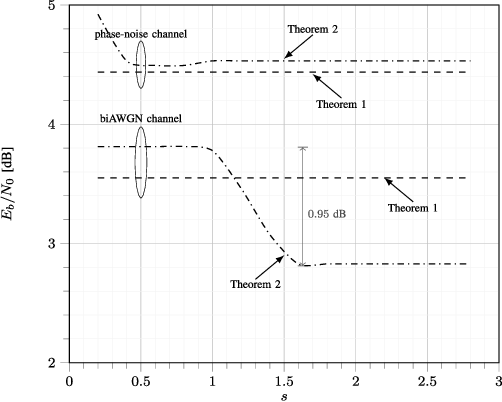}
	\caption{The minimum ${E_b}/{N_0}$ required to achieve $\pt^{\star}=10^{-3}$ and $\pu^{\star}=10^{-5}$ as a function of parameter $s$. Here, we consider $R = 0.5$ bits per channel use for the \ac{biAWGN} channel, $R = 1$ bits per channel use with 10 pilot symbols for the phase-noise channel; $k=50$ in both cases.}
	\label{fig:sVsSNR_ErrProb}
\end{figure}

We first observe that the gap between the two achievability bounds is largely reduced,
with the \ac{CRC}-based bound in Theorem~\ref{thm:Delta} actually yielding a slightly lower SNR threshold
than the threshold-based bound in Theorem~\ref{thm:Threshold_is} as the blocklength increases.
The much worse performance of the threshold-based decoding rule of Theorem~\ref{thm:Threshold_is}
compared to the \ac{biAWGN} case analyzed in Fig.~\ref{fig:ForneyBound} can be explained by the high
sensitivity to \ac{CSI} errors of this kind of decoders.
Furthermore, while for the
\ac{biAWGN} channel, significant performance improvements can be achieved by
optimizing the $s$ parameter in the bound, in the mismatched case such optimization
yields only very limited performance gains.
To clarify this point, we report in
Fig.~\ref{fig:sVsSNR_ErrProb}, the minimum \ac{SNR} required to achieve the target error
probability pair $\pt^* = 10^{-3}$, $\pu^* = 10^{-5}$ (as predicted by
Theorem~\ref{thm:Threshold_is}), as a function of the parameter $s$ for both \ac{biAWGN} and
block-memoryless phase-noise channel, with the phase estimated using $10$ pilot symbols. We
set $k=50$ for both cases. For comparison, we plot also the minimum \ac{SNR} predicted by
Theorem~\ref{thm:Delta}, which does not depend on $s$. We observe that for the \ac{biAWGN}
channel, optimizing over $s$ lowers the minimum required \ac{SNR} by $0.95\dB$. In
contrast, for the block-memoryless phase-noise channel, optimizing over $s$ results in a
negligible reduction in the minimum required \ac{SNR}.

Unlike threshold testing, when an outer \ac{CRC} code is used for error detection,
knowledge of the channel law has no impact on the trade-off between $\pt$ and $\pu$, as
this trade-off depends only on the number of \ac{CRC} bits, as shown in
\eqref{eq:ptDelta_def}--\eqref{eq:puDelta_def}. This observation is confirmed by the
simulation results reported in Fig.~\ref{fig:Mismatch}: Algorithm~\texttt{A}, which
relies on a \ac{CRC} code for error detection, largely outperforms Algorithm~\texttt{B}, which employs the threshold test~\eqref{eq:SCLtt}, at all blocklengths.
Furthermore, Algorithm~\texttt{B} outperforms the reference \ac{SCL} scheme only when $n=64$. This result suggests that, in the presence of inaccurate \ac{CSI}, the use
of a \ac{CRC}-based error detection
scheme should be preferred, even at very short blocklengths,
over thresholding-based schemes based on mismatched channel likelihoods.
Error detection by means of threshold testing is preferable only when sufficiently
accurate, albeit imperfect, \ac{CSI} is available at the decoder.

%%%%%%%%%%%%%%%%%%%%%%%%%%%%%%%%%%%%%%%%%%%%%%%%%%%%%%%%%%%%%%%%%%%%%%%%%%%%%%%%%%%%%%%%%%%%%%%%%%%%%%%%%%%%%%%%%%%%%%%%%%%%%%%%%%
\section{Conclusions}\label{sec:conclusions}
We analyzed the trade-off between the total error probability and the undetected error probability in
the short blocklength regime by presenting two finite-blocklength achievability bounds,
	which are used to benchmark the performance of CA polar codes.
The first finite-blocklength bound relies on a layered approach, where an outer code is
used to perform error detection. The second bound is based on a threshold test applied to
the generalized information density. On the \ac{biAWGN} channel and for blocklengths and error probabilities of interest in \ac{URLLC}, both bounds are more accurate than Forney's error-exponent achievability bound, with the bound based on generalized information density providing the best achievability result.

We also presented simulation results for CA polar codes
for three error-detection strategies:
\begin{inparaenum}[i)]
	\item the basic error-detection mechanisms provided by SCL decoding;
	\item an algorithm that splits the outer CRC parity bits in two subsets, with one subset dedicated to error detection;
	\item a threshold test that approximates Forney's optimal rule by exploiting the SCL decoder list.
\end{inparaenum}
Our results for the \ac{biAWGN} show that, in agreement with the finite-blocklength
	bounds, the threshold test applied at the output of the SCL decoder yields tangible
gains over the outer-\ac{CRC}-code approach. The gains decrease as the blocklength increases.

However, as illustrated for the case of block-memoryless phase-noise channel, the situation
	changes drastically when the decoder is fed with imperfect \ac{CSI}, which results in a
	mismatched decoding setting.
In this case, error
detection by means of an outer \ac{CRC} code exhibits a robust error detection capability,
whereas the use of threshold-based techniques that rely on a mismatched likelihood
results in  a significant performance loss for all blocklength values considered in the
paper.

\section*{Acknowledgment}
The authors would like to thank Prof. Gerhard Kramer (Technical University of Munich) for his constructive suggestions and the anonymous reviewers for their insightful comments that helped improve the final version of this paper.

%%%%%%%%%%%%%%%%%%%%%%%%%%%%%%%%%%%%%%%%%%%%%%%%%%%%%%%%%%%%%%%%%%%%%%%%%%%%%%%%%%%%%%%%%%%%%%%%%%%%%%%%%%%%%%%%%%%%%%%%%%%%%%%%%%
%%%%%%%%%%%%%%%%%%%%%%%%%%%%%%%%%%%%%%%%%%%%%%%%%%%%%%%%%%%%%%%%%%%%%%%%%%%%%%%%%%%%%%%%%%%%%%%%%%%%%%%%%%%%%%%%%%%%%%%%%%%%%%%%%%

\appendices
%%%%%%%%%%%%%%%%%%%%%%%%%%%%%%%%%%%%%%%%%%%%%%%%%%%%%%%%%%%%%%%%%%%%%%%%%%%%%%%%%%%%%%%%%%%%%%%%%%%%%%%%%%%%%%%%%%%%%%%%%%%%%%%%%%
\section{Proof of Theorem~\ref{thm:Delta}}\label{appendix:proof_Delta}
We consider the case when error detection is provided by an outer $(k+\Delta, k)$ code and error correction by an inner $(n, k+\Delta)$ code. Hence, $\Delta$ bits are used for error detection purposes.
We can think of the outer code as assigning each message $w\in\{1,\ldots,2^{k}\}$ to one of $2^{\Delta}$ bins. The inner encoder maps the $k$ bits describing the message, as well as the $\Delta$ bits describing the bin to which the message belongs, to a codeword of a codebook $\hat{\setC}$ with $|\hat{\setC}| = 2^{k+\Delta}$. Note that,
because of the bin assignment, only a subset $\setC \subseteq \hat{\setC}$ with $\abs{\setC} = 2^{k}$ of codewords is actually used.
This procedure allows us to select a subset $\setC$ of
cardinality $2^{k}$ from the codebook $\hat{\setC}$ of cardinality $2^{k+\Delta}$.
Let us denote by $\vecu \in \fieldtwo^{k}$ the binary representation of the
message to be transmitted, and by $\vecc\in \fieldtwo^{\Delta}$ the binary
representation of the corresponding bin index.
It will convenient to write $\vecc = b(\vecu)$ where the function $b(\cdot)$ returns the
binary representation of the index of the bin in which $\vecu$ is placed.

The decoder first selects a codeword in $\hat{\setC}$ uniformly at random among all codewords that have maximum likelihood. If this codeword belongs also to $\setC$, then the corresponding message is returned. If not, the decoder declares an erasure.

Next, we evaluate the average total and undetected error probabilities that are achievable by this coding scheme. Specifically, we average over both the codebook, which is generated by drawing each codeword independently from $P_{\vecx}$, and the assignment of messages to bins, which we assume is performed uniformly at random.

Let $\hat{\setC} = \{\vecx_{1},\ldots,\vecx_{2^{k+\Delta}}\}$.
Let also assume, without loss of generality, that the pair $(\vecu,\vecc)$
associated to the message to be transmitted result in the encoder selecting
codeword $\vecx_1$.
For $m = 2,\ldots,2^{k+\Delta}$, we let
\begin{equation}
	\label{eq:psiDef}
	\psi_{\rvecy,\rvecx}(\vecy,\vecx_1) =  \Prob\ltrsqr{P_{\rvecy|\rvecx}(\vecy|\rvecx_m) \geq P_{\rvecy|\rvecx}(\rvecy|\rvecx_1) \ggiven \rvecx_1,\rvecy }.
\end{equation}
Note that this quantity is the same for all $m$ since the codewords $\rvecx_{m}$ are identically distributed.
An achievability bound on $\pt$ follows directly by applying the RCU bound~\cite[Thm.~16]{Polyanskiy2010} to the codebook $\hat{\setC}$.
Specifically,
\begin{align}
	\label{eq:DeltaRCU_1}
	\pt & \leq \Prob\ltrsqr{ \Union_{m=2}^{2^{k+\Delta}} P_{\rvecy|\rvecx}(\rvecy|\rvecx_{m}) \geq P_{\rvecy|\rvecx}(\rvecy|\rvecx_1)}                                                       \\
	    & = \Ex{}{ \Prob\ltrsqr{ \Union_{m=2}^{2^{k+\Delta}} P_{\rvecy|\rvecx}(\rvecy|\rvecx_{m}) \geq P_{\rvecy|\rvecx}(\rvecy|\rvecx_1) \ggiven \rvecx_1,\rvecy} } \label{eq:DeltaRCU_2.2} \\
	    & \leq \Ex{}{\min\ltrcurley{1, \sum_{m=2}^{2^{k+\Delta}} \psi_{\rvecy,\rvecx}(\vecy,\vecx_1)}} \label{eq:DeltaRCU_2.9}                                                               \\
	    & = \mathrm{RCU}(k+\Delta,n) \label{eq:DeltaRCU_3}.
\end{align}
Here, the upper bound in \eqref{eq:DeltaRCU_1} follows by assuming that an error always occurs if more than one codeword has maximum likelihood, \eqref{eq:DeltaRCU_2.9} follows from the union bound, and~\eqref{eq:DeltaRCU_3} follows  by letting $\rvecx$ be the transmitted codeword and $\bar{\rvecx}$ be one of the other codewords.
To upper-bound $\pu$, we proceed as follows: let $\hat{m}=\argmax_{m \in \{1,\dots, 2^{k+\Delta}\}}
	P_{\rvecy\given\rvecx}(\rvecy\given \rvecx_{m})$ and let the pair $(\hat{\vecu},
	\hat{\vecc})$ denote the binary representation of the index $\hat{m}$, where $\hat{\vecu}$
corresponds to the estimate of the binary representation of the transmitted message, and
$\hat{\vecc}$ is the estimate of the binary representation of its bin index.
An error is undetected if $\hat{m}\neq 1$ and $b(\hat{\vecu})=\hat{\vecc}$.
Hence,
\begin{IEEEeqnarray}{rCl}
	\pu
	&=& \Prob[\hat{m}\neq 1, b(\hat{\vecu})=\hat{\vecc}]\\
	&=& \Prob[\hat{m}\neq 1] \Prob[b(\hat{\vecu})=\hat{\vecc} \given \hat{m}\neq 1]\\
	&\leq&  \Prob\ltrsqr{ \Union_{m=2}^{2^{k+\Delta}} P_{\rvecy|\rvecx}(\rvecy|\rvecx_{m})\geq P_{\rvecy|\rvecx}(\rvecy|\rvecx_1)} 2^{-\Delta}\label{eq:DeltaRCU_pu2}\\
	&\leq&\mathrm{RCU}(k+\Delta,n)2^{-\Delta}\label{eq:DeltaRCU_pu4}.
\end{IEEEeqnarray}
Here, \eqref{eq:DeltaRCU_pu2} follows by using the same bound as in \eqref{eq:DeltaRCU_1}; we also used that the probability that the message with binary representation $\hat{\vecu}$ is assigned to the bin with binary representation $\hat{\vecc}$ is $2^{-\Delta}$. Finally,~\eqref{eq:DeltaRCU_pu4} follows from the same steps leading to~\eqref{eq:DeltaRCU_3}.
\QEDA

%%%%%%%%%%%%%%%%%%%%%%%%%%%%%%%%%%%%%%%%%%%%%%%%%%%%%%%%%%%%%%%%%%%%%%%%%%%%%%%%%%%%%%%%%%%%%%%%%%%%%%%%%%%%%%%%%%%%%%%%%%%%%%%%%%
\section{Proof of Theorem~\ref{thm:Threshold_is}}\label{appendix:proof_threshold_is}
We consider a decoder that outputs the message that corresponds to the codeword that is selected uniformly at random among the codewords with the largest likelihood $P_{\rvecy|\rvecx}(\vecy|\vecx)$, provided that its generalized information density, defined in~\eqref{eq:GenInfoDens}, exceeds the threshold $n\lambda$; if not, the decoder declares an erasure.
We next use a random coding argument and evaluate the average \ac{TEP}, averaged over random codebooks constructed by drawing each codeword independently from $P_{\rvecx}$. Let us assume, without loss of generality, that $\vecx_{1}$ is the transmitted codeword. Then,
\begin{align}
	\pt & \leq \Prob\mathopen{}\vast[\ltrcurley{\infden(\rvecx_1,\rvecy) < n \lambda } \Union \nonumber                                                                             \\
	    & \quad \ltrcurley{\infden(\rvecx_1,\rvecy) \geq n \lambda , \Union_{m=2}^{2^{k}} P_{\rvecy|\rvecx}(\rvecy|\rvecx_m) \geq P_{\rvecy|\rvecx}(\rvecy|\rvecx_1) } \vast]       \\
	    & \leq \Prob\ltrsqr{\infden(\rvecx_1,\rvecy) \geq n \lambda , \Union_{m=2}^{2^{k}} P_{\rvecy|\rvecx}(\rvecy|\rvecx_m) \geq P_{\rvecy|\rvecx}(\rvecy|\rvecx_1)} \nonumber    \\
	    & \quad + \Prob\ltrsqr{\infden(\rvecx_1,\rvecy) < n \lambda} \label{eq:RCU_Thr_2}                                                                                           \\
	    & =  \Exop\mathopen{}\vast[\Prob \ltrsqr{ \Union_{m=2}^{2^{k}} P_{\rvecy|\rvecx}(\rvecy|\rvecx_m) \geq P_{\rvecy|\rvecx}(\rvecy|\rvecx_1)\Bigg| \rvecx_1,\rvecy}  \nonumber \\
	    & \quad \times \ind{\infden(\rvecx_1,\rvecy) \geq n \lambda}  \vast] + \Prob\ltrsqr{\infden(\rvecx_1,\rvecy) < n \lambda}                                                   \\
	    & \leq  \Ex{}{\min\ltrcurley{1, \sum_{m=2}^{2^{k}} \psi_{\rvecy,\rvecx}(\vecy,\vecx_1)} \ind{\infden(\rvecx_1,\rvecy) \geq n \lambda}} \nonumber                            \\
	    & \quad  + \Prob\ltrsqr{\infden(\rvecx_1,\rvecy) < n \lambda} \label{eq:RCU_Thr_4}                                                                                          \\
	    & \leq \widetilde{\mathrm{RCU}}_{\lambda}(k,n) + \Prob\ltrsqr{\infden(\rvecx,\rvecy) < n \lambda} \label{eq:RCU_Thr_5}.
\end{align}
Here, \eqref{eq:RCU_Thr_2} and \eqref{eq:RCU_Thr_4} follow from the union bound (note that $\psi_{\vecy,\vecx}(\vecy,\vecx_1)$ was defined in \eqref{eq:psiDef}); \eqref{eq:RCU_Thr_5} follows by letting $\rvecx$ be the transmitted codeword and $\bar{\rvecx}$ be one of the other codewords.

Similarly, we can bound $\pu$ as
\begin{align}
	\pu & \leq \Prob\ltrsqr{\Union_{m=2}^{2^{k}} P_{\rvecy|\rvecx}(\rvecy|\rvecx_{m}) \geq \max\ltrcurley{P_{\rvecy|\rvecx}(\rvecy|\rvecx_1), \widetilde{\lambda}_{\vecy} } } \label{eq:ThrPU1}                        \\
	    & \leq \Exop\mathopen{}\bigg[\min\bigg\{1, (2^{k}-1)  \nonumber                                                                                                                                                \\
	    & \quad \times \Prob\ltrsqr{P_{\rvecy|\rvecx}(\rvecy|\bar{\rvecx}) \geq \max\ltrcurley{P_{\rvecy|\rvecx}(\rvecy|\rvecx), \widetilde{\lambda}_{\vecy} } \ggiven \rvecx, \rvecy}\bigg\} \bigg].\label{eq:ThrPU2}
\end{align}
Here, \eqref{eq:ThrPU1} follows because the condition $\infden(\vecx_m,\rvecy) \geq n\lambda$ is equivalent to $P_{\rvecy|\rvecx}(\rvecy|\rvecx_m) \geq \widetilde{\lambda}_{\vecy}$; \eqref{eq:ThrPU2} follows from steps similar to the ones leading to \eqref{eq:RCU_Thr_5}.

With these steps, we have established achievability bounds on the average \ac{UEP} and
\ac{TEP}, averaged over the random codebook.
However, this does not imply the existence of a single deterministic
codebook that achieves both bounds simultaneously.
To solve this problem, we proceed as in~\cite[App.~A]{Johan2021} (see also~\cite[Thm.~3]{Polyanskiy2011}),
and conclude that the bounds can be achieved by a randomized coding strategy that
involves time sharing between two codebooks. Time-sharing is made possible by the introduction of the random variable $u$ (see Definition~\ref{def:code}).\QEDA


%%%%%%%%%%%%%%%%%%%%%%%%%%%%%%%%%%%%%%%%%%%%%%%%%%%%%%%%%%%%%%%%%%%%%%%%%%%%%%%%%%%%%%%%%%%%%%%%%%%%%%%%%%%%%%%%%%%%%%%%%%%%%%%%%%

\end{document}